\documentclass[aps,prb, twocolumn, groupedaddress, showpacs]{revtex4}
\usepackage{amsmath}
\usepackage{amssymb}
\usepackage{graphicx}
\usepackage{tikz}
\begin{document}

\title{Transport in line junctions of $\nu=5/2$ quantum Hall liquids}
\author{Chenjie~Wang and D.~E.~Feldman}
\affiliation{Physics Department, Brown University, Providence, Rhode
Island 02912, USA}

\date{\today}

\begin{abstract}
We calculate the tunneling current through long line junctions of a $\nu=5/2$ quantum Hall liquid and i) another $\nu=5/2$ liquid, ii) an integer quantum Hall liquid and iii) a quantum wire. Momentum resolved tunneling provides information about the number, propagation directions and other features of the edge modes and thus helps distinguish several competing models of the 5/2 state. We investigate transport properties of two proposed Abelian states: $K=8$ state and 331 state, and four possible non-Abelian states: Pfaffian, edge-reconstructed Pfaffian, and two versions of the anti-Pfaffian state. We also show that the non-equilibrated anti-Pfaffian state has a different resistance from other proposed states in the bar geometry.
\end{abstract}


\maketitle

\section{Introduction}

\begin{figure}
\centering
\includegraphics[width = 3.3in]{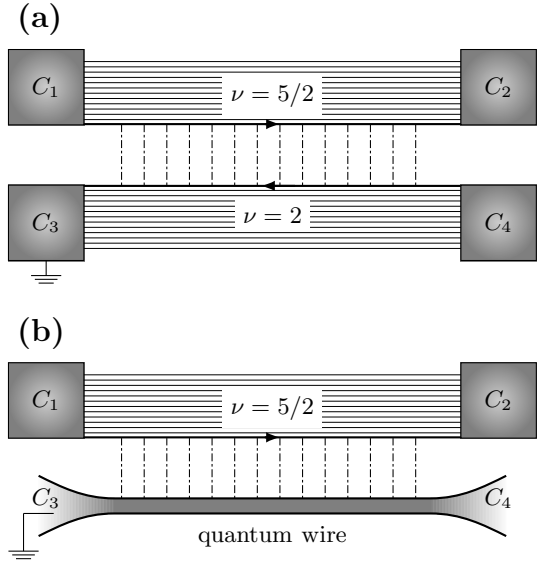}\caption{(a) Tunneling between $\nu=5/2$ and $\nu=2$ QHE liquids. The edges of the upper
and lower QHE liquids form a line
junction. (b) Tunneling between $\nu=5/2$ QHE liquid and a quantum wire. In both setups, contacts $C_1$ and $C_2$ are kept at the same voltage $V$.} \label{fig:setup}
\end{figure}

One of the most interesting aspects of the quantum Hall effect (QHE)
is the presence of anyons which carry fractional charges and
obey fractional statistics. In many quantum Hall states,
elementary excitations are Abelian anyons \cite{prange-book87}. They accumulate non-trivial statistical
phases when move around other anyons and
can be viewed as charged particles with infinitely long solenoids attached.
A more interesting theoretical possibility involves non-Abelian anyons \cite{nayak08}. In contrast to Abelian QHE states,
non-Abelian systems change not only
their wave functions but also their quantum states when one anyon encircles another.
This property makes non-Abelian anyons a promising tool for quantum information processing \cite{kitaev03}.
However, their existence in nature remains an open question.

It has been proposed that non-Abelian anyons might exist in the QHE liquid at the filling factor $\nu=5/2$, Ref.~\onlinecite{moore91}.
Possible non-Abelian states include different versions of Pfaffian and anti-Pfaffian states \cite{lee07,levin07,overbosch08b}.
At the same time, Abelian candidate wave functions such as $K=8$ and 331 states were also suggested \cite{overbosch08b, wen92b} for $\nu=5/2$. Different models predict different quasiparticle statistics but the same quasiparticle charge
$q=e/4$, where $e<0$ is an electron charge. Since the experiments \cite{dolev08,radu08,willett09} have been limited to the determination of the charge of the elementary excitations,
the correct physical state remains unknown.

Several methods to probe the statistics in the 5/2 state were suggested but neither was successfully implemented so far. This motivates
further investigations of possible ways to test the statistics.
The definition of exchange statistics involves quasiparticle braiding. Hence, interferometry is a natural choice.
An elegant and conceptually simplest interferometry
approach involves an anyonic Fabry-Perot interferometer \cite{chamon97, fradkin98, dassarma05, stern06, bonderson06}. Its practical implementation faces difficulties due in part to the fluctuations of the trapped topological charge \cite{kane03,fluct09}.
A very recent Fabry-Perot experiment might have shown a signature of anyonic statistics \cite{wpw}. However, interpretation of such experiments
is difficult \cite{nayak-int} and must take into account sample-specific factors such as Coulomb blockade effects. \cite{hr-cb,marcus-cb}
An approach based on a Mach-Zehnder interferometer \cite{ji03,  law06, feldman06, feldman07, law08} is not sensitive to slow fluctuations of the trapped topological charge but just like the Fabry-Perot interferometry it cannot easily distinguish Pfaffian and anti-Pfaffian states. On the other hand, the structure of edge states contains full information about the bulk quantum Hall liquid and thus a tunneling experiment with a single quantum point contact might be sufficient
\cite{overbosch08b}. Unfortunately, even in the case of simpler Laughlin states the theory and experiment have not been reconciled for this type of measurements\cite{chang03}. Besides,
the scaling behavior of the tunneling $I-V$ curve is non-universal and depends on many factors such as edge reconstruction \cite{edge-rec94} and  long range Coulomb interactions. An approach based on two-point-contact geometry \cite{feldman08} identifies different states through their universal signatures in electric transport. This comes at the expense of the
necessity to measure both current and noise. Recently an approach based on tunneling through a long narrow strip of the quantum Hall liquid was proposed\cite{overbosch08a}.
This approach, however, has the same limitation as the Fabry-Perot geometry: interference is smeared by the quasiparticle tunneling into and from the strip. In this paper we analyze a related approach with tunneling through a long narrow line junction of quantum Hall liquids and a line junction of a $\nu=5/2$ quantum Hall liquid and a quantum wire. Since only electrons tunnel in such geometry, the interference picture is not destroyed by quantum fluctuations.

Fig.~\ref{fig:setup} shows sketches of our setups. Electrons tunnel from the $\nu = 5/2$ fractional QHE state to  the $\nu = 2$ or $\nu=1$ integer QHE state through a line junction in the weak tunneling regime (Fig.~\ref{fig:setup}a)) at near zero temperature. A similar setup has already been realized in the integer QHE regime\cite{kang00}.
Fig.~\ref{fig:setup}b) illustrates a setup with electron tunneling between the
edge of the $\nu=5/2$ liquid and a one-channel quantum wire.
The most important
feature in these setups is the conservation of both energy and
momentum in each tunneling event\cite{kang00,zulicke03,melikidze04}. The two conservation laws lead to singularities in the $I-V$ curve. Each singularity emerges due to one of the edge modes on one side of the junction. Thus, the setups allow one to count the modes and distinguish different proposed states since they possess different numbers
and types of edge modes with different propagation directions and velocities.
In particular, these setups are able to distinguish different Abelian and non-Abelian states.

\begin{figure}
\centering
\includegraphics[width=3in]{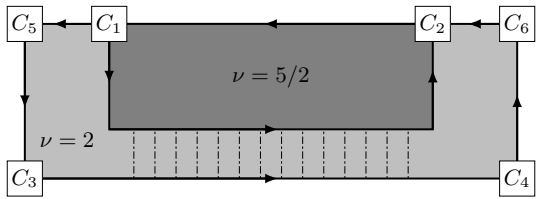}
\caption{Tunneling between the fractional QHE channels of the $\nu=5/2$ edge  and the $\nu=2$ integer channels.  Contacts $C_1$ and $C_2$ are kept at the same voltage $V$ and the other contacts are grounded. } \label{setup-2}
\end{figure}

The edge of the 5/2 state includes both a fractional 5/2 edge and two integer quantum Hall channels.
In the setups Fig.~\ref{fig:setup},  electrons tunnel both into the fractional and integer channels on the edge. However, our calculations are also relevant for a setup in which tunneling occurs into the fractional 5/2 edge only. Such situation can be achieved in the way illustrated in
Fig.~\ref{setup-2},
similar to experiments \cite{deviatov-2002, deviatov-2006, deviatov-2008}. In the setup Fig.~\ref{setup-2}, a voltage difference is created between integer and fractional quantum Hall channels on the same edge and tunneling occurs between the integer and fractional channels.
Our results also apply to the setup considered in Ref. \onlinecite{seidel09}. In that setup, tunneling occurs into an edge separating $\nu=2$ and $\nu=5/2$
quantum Hall liquids.

The paper is organized as follows. We
review several models of the $5/2$ state and their
corresponding edge modes in Sec.~\ref{sec:states}.
Sec.~\ref{sec:qualitative} contains a qualitative discussion of the momentum resolved tunneling. We describe our technical approach in Sec.~\ref{sec:calculations}. The number of conductance singularities allows one to distinguish different models. This number is computed in section ~\ref{sec:V}. Detailed calculations of the $I-V$ curve for each edge state are given in Sec.~\ref{sec:results}
in the limit of weak interactions between fractional and integer edge channels. Our results are summarized in Sec.~\ref{sec:discussion}. We discuss
effects of possible reconstruction of integer QHE modes in the Appendix.

\section{Proposed $5/2$ states}
\label{sec:states}

Numerical experiments\cite{morf98, rezayi00, moller, DMRG} generally support a spin-polarized state for the quantum Hall liquid with $\nu=5/2$.
Below we review the simplest spin-polarized candidate states, including the abelian $K
= 8$ state, a version of the $331$ state, and non-abelian Pfaffian and anti-Pfaffian states. In all those states, the lowest Landau level is fully
filled with both spin-up and spin-down electrons which form two
integer QHE liquids, while in the second Landau level electrons form
a spin-polarized $\nu = 1/2$ fractional QHE liquid.
Our approach can be easily extended to spin-unpolarized states.
In the following, we focus on the $1/2$ fractional QHE liquid and its edge.
The lowest Landau level contributes two more edge channels.

The $K=8$ state can be understood as a quantum Hall state of Cooper pairs.
The $331$ state is formed by the condensation of the charge-$2e/3$ quasiparticles on top of the Laughlin $\nu=1/3$ state. A different version
of the $331$ state is also known \cite{haldane88}. Since that version is not spin-polarized, we do not consider it below.

The abelian $K = 8$ and $331$ states \cite{overbosch08b} can be described by
Ginzburg-Landau-Chern-Simons effective theories\cite{wen92a}, with the
Lagrangian density given by
\begin{equation}
{\cal L} = -\frac{\hbar}{4\pi}\sum_{IJ\mu\nu} K_{IJ} a_{I\mu} \partial_{\nu}
a_{J\lambda}\epsilon^{\mu\nu\lambda},
\end{equation}
where $\mu,\nu=t,x,y$ are space-time indices.
The $K$-matrix describes the topological orders of the bulk, and its
dimension gives the number of layers in the hierarchy. The $U(1)$
gauge field $a_{I\mu}$  describes the quasiparticle/quasihole density
and current in the $I$th hierarchical condensate. This effective
bulk theory also determines the theory at the edge, where the $U(1)$
gauge transformations are restricted. The edge theory, called chiral
Luttinger liquid theory, has the Lagrangian density
\begin{equation}
{\cal L}_{\rm edge} = -\frac{\hbar}{4\pi}\sum_{IJ}(\partial_t\phi_I K_{IJ}
\partial_x\phi_J + \partial_x\phi_I V_{IJ}
\partial_x\phi_J).
\end{equation}
The chiral boson field $\phi_I$ describes gapless edge excitations
of the $I$th condensate, and $V_{IJ}$ is the interaction between the
edge modes. We see that the dimension of the $K$-matrix gives
the number of the edge modes. In the $K = 8$ state, electrons first pair
into charge-$2e$ bosons, then these bosons condense into a $\nu =
1/8$ Laughlin state. Hence, the $K$-matrix is a $1\times1$
matrix whose only element equals 8, and so there is only one right-moving edge mode. The $331$ state is
characterized by
\begin{equation}
K = \left(
\begin{array}{cc}
3& -2\\
-2& 4
\end{array}\right)
\end{equation}
which has two positive eigenvalues, so there are two right-moving modes at the
edge. This state should be contrasted with the spin-unpolarized version of the 331 state, whose $K$-matrix has entries equal to 3 and 1 only.
The same name is used for the two states since they have the same topological order \cite{overbosch08b}.

The Pfaffian state \cite{moore91} can be described by the following wave function for the $1/2$ fractional QHE liquid
\begin{equation}
\Psi_{\rm Pf} = \text{Pf}(\frac{1}{z_i - z_j})\prod_{i<j}(z_i -
z_j)^2 e^{-\sum_i |z_i|^2},
\end{equation}
in which $z_n=x_n+iy_n$ is the coordinate of the $n$th electron in units of
the magnetic length $l_B$, and Pf is the Pfaffian of the
antisymmetric matrix $1/(z_i - z_j)$. At the edge, there
is one right-moving charged boson mode and one right-moving neutral
Majorana fermion mode. The edge action assumes the form (\ref{pfedgeaction}). In the presence of edge reconstruction, the action changes\cite{overbosch08b}. In the reconstructed edge state, there are one right-moving charged and one right-moving neutral boson mode, and one left-moving neutral Majorana fermion mode. The edge action becomes Eq.~(\ref{recpfedgeaction}).

The anti-Pfaffian state \cite{lee07,levin07} is the particle-hole conjugate of the Pfaffian state, i.e.,
the wave function of the anti-Pfaffian state can be obtained from the Pfaffian wave function through a particle-hole
transformation\cite{girvin84}, given in Ref.~\onlinecite{lee07}.
There are two versions of the anti-Pfaffian edge states. One possibility is a
non-equilibrated edge. In that case tunneling between different edge modes can be neglected
and the modes do not equilibrate. The action contains two
counter-propagating charged boson modes and one left-moving neutral
Majorana fermion mode Eq.~(\ref{noneqaction}). The other version is the disorder-dominated state, in
which there are one right-moving charged boson mode and three
left-moving neutral Majorana fermion modes of exactly the same
velocity, Eq.~(\ref{eqaction}). As discussed below, only limited information about the latter state can be extracted from the transport through a line junction since momentum does not conserve in tunneling to a disordered edge.

We see from the above discussion that different proposed edge states have
different numbers and types of modes. This important information can be used to detect the nature of the 5/2 state as discussed in the rest of
this paper.


\section{Qualitative discussion}
\label{sec:qualitative}

In this section we discuss some details of the setup.
We also provide a qualitative explanation of the results of the subsequent sections in terms
of kinematic constraints imposed by the conservation laws.

Our setups are shown in  Fig.~\ref{fig:setup}. The long uniform junction couples the edge of
the upper $\nu = 5/2$ fractional QHE liquid with the edge of the lower
$\nu = 2$ or $\nu=1$ integer QHE liquid. Such a system with two
sides of the junction having different filling factors can be
realized experimentally in semiconductor heterostructures with two mutually
perpendicular 2D electron gases (2DEG)\cite{huber02,zulicke03}.
Properly adjusting the direction and magnitude of the magnetic field
one can get the desired filling factors\cite{zulicke03}.
Depending on the direction of the magnetic field, the upper and lower
edge modes in Fig. 1a) can be either co- or counter-propagating.
In Sec.~\ref{sec:discussion}, we will  also briefly discuss the tunneling between two
5/2 states. This situation can be realized by introducing a barrier
in a single 2DEG \cite{kang00}. We will
see however that the second setup is less informative than the first one.
Finally, we will consider tunneling between a 5/2 edge and a uniform parallel one-channel quantum wire.
Such setup can come in two versions: a) tunneling into a full 5/2 edge that includes both fractional and integer modes
and b) tunneling into a fractional edge between $\nu=2$ and $\nu=5/2$ QHE liquids \cite{seidel09}. A closely related setup
is illustrated in Fig. \ref{setup-2}. There the tunneling occurs between different modes of the same edge.

Below we will use the language referring to tunneling between two QHE liquids, a 5/2 liquid and an integer $\nu=2$ QHE liquid.
This language can be easily translated to the quantum wire situation. In contrast to
the integer QHE edge, a quantum wire contains counter-propagating  modes. However, the energy and momentum conservation, together with the Pauli principle,
generally restrict tunneling to only one of those modes.

The Hamiltonian assumes the following general structure:
\begin{equation}
\label{Ham-general}
H=H_{5/2}+H_{\rm int}+H_{\rm tun},
\end{equation}
where the three contributions denote the Hamiltonians of the $5/2$ edge, the integer edge and the tunneling term.
The latter term expresses as
\begin{equation}
\label{Ham-tunneling}
H_{\rm tun}=\int dx \psi^\dagger(x)\sum_n \Gamma_n(x)\psi_n(x)+ {\rm H.c.} ,
\end{equation}
where $x$ is the coordinate on the edge, $\psi^\dagger(x)$ is the electron creation operator at the integer edge,
$\psi_n$ are electron operators at the fractional QHE edge and $\Gamma_n(x)$ are tunneling amplitudes.
Several operators $\psi_n$ correspond to different edge modes.
We assume that the system is uniform. This imposes a restriction
\begin{equation}
\label{restriction}
\Gamma_n(x)\sim\exp(-i\Delta k_n x),
\end{equation}
where $\Delta k_n$ should be understood as the momentum mismatch between different modes. In order to derive Eq.~(\ref{restriction}) we
first note that in a uniform system $|\Gamma_m(x)|$ cannot depend on the coordinate.
Next, we consider the system with the tunneling Hamiltonian $H_{\rm tun}'=\psi^\dagger(x_0)\Gamma_m(x_0)\psi_m(x_0)+\psi^\dagger(x_0+a)\Gamma_m(x_0+a)\psi_m(x_0+a)
+{\rm H.c. }$  The current can depend on $a$ only and not on $x_0$ - otherwise different points of the junction would not be equivalent.
Applying the second order perturbation theory in $\Gamma_m$ to the calculation of the current one finds that $\Gamma_m(x_0)\Gamma^*_m(x_0+a)$ must be a constant, independent of $x_0$. Using the limit of small $a$ one now easily sees that the phase of the complex number $\Gamma_m(x)$ is a linear function of $x$. This proves Eq.~(\ref{restriction}).

We assume that the same voltage $V$ is applied to both contacts at the upper $\nu=5/2$ edge in Fig.~1, so
that all right-moving and left-moving modes at the upper edge are in
equilibrium with the chemical potential $\mu_1=eV$. The lower edge is grounded, i.e., the chemical potential at the lower edge $\mu_2 =
0$.

$\Delta k_n$ may depend on the applied voltage $V$ since the
width of the line junction may change when the applied voltage
changes. We will neglect that dependence in the case of the setup with the tunneling between two QHE liquids;
more specifically,
we will assume that both liquids are kept at a constant charge density and the tunneling between them is weak.
In the case of the tunneling between a QHE liquid and a quantum wire we will assume that the charge density is kept constant in 2DEG
but can be controlled by the gate voltage in the one-dimensional wire. The Fermi-momentum $k_F$ in the quantum wire depends on the charge density and
any change of $k_F$ results in an equal change of all $\Delta k_n$.
Thus, we will assume a setup with two 2DEG in the discussion of the voltage dependence of the tunneling current at fixed $\Delta k_n$.
The setup with a quantum wire will be assumed in the discussion of the dependence of the current on $k_F$ at a fixed low voltage. In all cases we will assume that the temperature is low.

In our calculations we will use the Luttinger liquid model for the edge states \cite{wen-book04}. It assumes a linear spectrum for each mode and neglects tunneling
between different modes on the same edge. These assumptions are justified in the regime of low energy and momentum.
Thus, we expect that the results for the tunneling between two 2DEG are only qualitatively valid at high voltage.

\begin{figure}
\centering
\includegraphics[width = 3.3in]{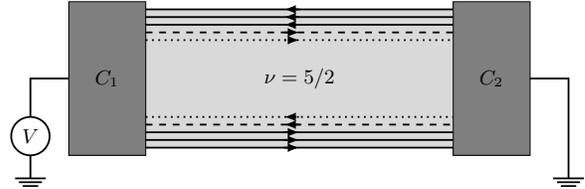}\caption{A bar geometry that can be used to detect the non-equilibrated anti-Pfaffian state. Solid lines denote Integer QHE edge modes, the dashed lines denote fractional QHE charged modes and dotted lines denote Majorana modes. Arrows show mode propagation directions. } \label{fig:setup2}
\end{figure}

Our main assumption is that both energy and momentum conserve in each tunneling event. This means that we neglect disorder at the edges. This assumption needs a clarification in the case of the disorder-dominated anti-Pfaffian state because its formation requires edge disorder. We will assume that for that state only neutral modes couple to disorder and one can neglect disorder effects on the charged mode. For completeness, we include a discussion of the momentum resolved tunneling into the non-equilibrated anti-Pfaffian state. However, a much simpler experiment is sufficient
to detect that state. One just needs to measure the conductance of the $5/2$-liquid in the bar geometry illustrated in Fig.~\ref{fig:setup2}. Indeed, in the non-equilibrated anti-Pfaffian state, disorder is irrelevant. Each non-equilibrated edge has
three charged Fermi-liquid modes propagating in one direction and another Luttinger-liquid charged mode (and a neutral mode) propagating in the opposite direction. In the bar geometry, the lower edge carries the current $3e^2V/h$. The upper charged mode carries the current $e^2V/(2h)$ {\it in the same direction}. Hence, the total current is $7e^2V/(2h)$ and the conductance is $7/2$ and not $5/2$ conductance quanta. Our discussion assumes an ideal situation with no disorder. In a large system even weak disorder, irrelevant in the renormalization group sense, might result in edge equilibration. Nevertheless, if the QHE bar is shorter than the equilibration length the nature of the state can be probed by the conductance measurement in the bar geometry.

Before presenting the calculations we will discuss a qualitative picture. Unless otherwise specified we consider $\Delta k_n>0$. As seen from the calculations in the following section, the particle-hole symmetry for Luttinger liquids implies that the tunneling current at negative $\Delta k_n$
can be found from the relation $I_{\rm tun}(V, \Delta k )= - I_{\rm tun}(-V, -\Delta k)$. We assume that tunneling is weak and hence only single electron tunneling matters. One can imagine two types of electron operators on the edge: one type of operators simply creates an electron in one of the integer or fractional channels. The second type of operators creates and destroys electrons in different edge channels of the same edge. Generally, operators of the second type are less relevant than operators of the first type and we will neglect them (see, however, a discussion in the Appendix for the case of reconstructed integer edge channels). An exception is the $K=8$ state. Only electrons pairs can tunnel into the fractional $K=8$ edge.
As we will see in section VI, the most relevant single-electron operator
transfers two electron charges into the fractional edge and removes one electron charge from a co-propagating integer edge. For simplicity of our qualitative discussion, in this section we will disregard that operator and concentrate instead on the two-electron tunneling operator into the fractional edge. Such operator is most relevant in the setup Fig \ref{setup-2}.

We will use another simplifying assumption in this section: we will neglect interaction between different integer and fractional modes.
This assumption is not crucial as discussed in section V and we make it solely for simplicity. We will find the total number of singularities both for strongly and weakly interacting edges. At the same time, the current can be found analytically in the case of weak interactions, Sec. VI.

\begin{figure}
\centering
\includegraphics[width = 3.3in]{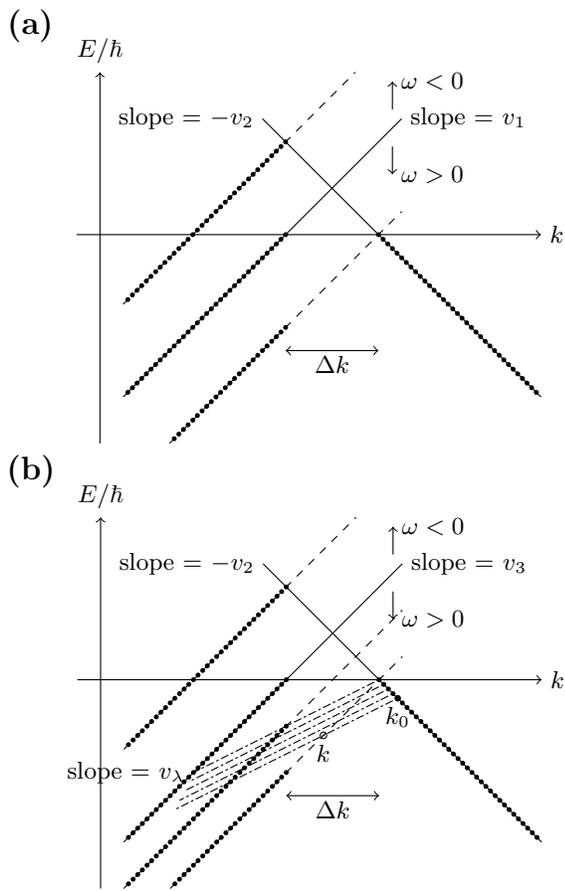}
\caption{Illustration of the graphical method.
(a) Tunneling between two integer QHE modes.
The left solid line represents the electron
spectrum at the upper edge at zero voltage. The right solid line represents the spectrum at the lower edge. The dashed lines represent the electron spectra at the upper edge
at different voltages. Black dots represent occupied states. The momentum mismatch between two edges
$\Delta k>0$. (b) Tunneling between an integer QHE edge and a Pfaffian edge. The right line represents the spectrum of the integer edge. The left line shows the spectrum of the charged boson mode at the Pfaffian edge. The unevenly dashed lines ($\lambda$ lines) represent Majorana fermions. The figure illustrates a tunneling event in which an electron with the momentum $k_0$ tunnels into the Pfaffian edge and creates a boson with the momentum $k$ and a Majorana fermion with the momentum $k_0-k$.}\label{fig:conservation}
\end{figure}

At the lower edge there are two edge modes for spin-up and -down electrons. At the upper edge there are two spin-up and -down integer modes
and one or more modes corresponding to the $\nu=1/2$ edge. Spin is conserved during the tunneling process. Thus, we have three contributions to the tunneling current: (A) tunneling between the upper spin-down fractional edge modes and the lower spin-down  integer edge mode; (B) tunneling between the upper spin-down integer edge mode and the lower spin-down integer edge mode; (C) tunneling between the upper spin-up integer edge mode and the lower spin-up integer edge mode. We use only the lowest order perturbation approximation so these contributions are independent. Thus, the total tunneling current is  $I_{\rm tun}=I_{\rm tun}^A+I_{\rm tun}^B+I_{\rm tun}^C$.  Contributions (B) and (C) are similar since the Zeeman energy is small compared to the Coulomb interaction under typical magnetic fields. Thus, we will only consider spin-down electrons below.

All edge modes are chiral Luttinger liquids with the spectra of the form $E=\pm v_\alpha(k-k_{F\alpha})$, where $\pm v_\alpha$ is the edge mode
velocity,
the sign reflects the propagation direction. We will first consider case (B) (case (C) is identical), tunneling between two integer Fermi-liquid edge modes.
Denote the upper edge velocity as $v_1$ and the lower edge velocity as $v_2$. If an electron of momentum $k$ from the upper edge tunnels into the lower edge or vice verse, energy and momentum conservation gives \begin{align}
v_1(k-k_{F1}) - \omega &= -v_2(k-k_{F2}),
\label{eqn:conservation}
\end{align}
where $\omega = -eV/\hbar$ (in the rest of this paper, we will refer to both $\omega$ and $V$ as the applied voltage).
The tunneling happens only when $(k-k_{F1})(k-k_{F2})<0$, i.e., one of the two states is occupied and the other is not. Eq.~(\ref{eqn:conservation}) is easy to solve directly but a graphical approach is more transparent.
Fig.~\ref{fig:conservation}(a) shows the spectra in the energy-momentum space, where the left line describes the upper edge mode and the right line describes the lower edge mode, and the intersection point represents the solution of Eq.~(\ref{eqn:conservation}). The black dots represent occupied states. We see that when $\omega = 0$, both states at the intersection point are unoccupied, therefore no tunneling happens. When $\omega$ increases, the left line moves down. For a small $\omega$, there is still no tunneling. After $\omega$ reaches the value of $v_1\Delta k= v_1(k_{F2}-k_{F1})$ and  the state from the right line at the intersection point becomes occupied, an electron from the lower edge can tunnel into the upper edge. This results in  a positive contribution to the tunneling current. Since the tunneling happens only at the intersection point and the tunneling density of states (TDOS) is a constant in Fermi liquids, the current will remain constant for $\omega>v_1\Delta k$. For a negative $\omega$, the situation is similar. Before $\omega$ reaches the value $-v_2\Delta k$, i.e., $|\omega|<v_2\Delta k$, no tunneling happens. When $|\omega|>v_2\Delta k$, an electron from the upper edge can tunnel into the lower edge and a negative voltage-independent tunneling current results. Thus, the $I_{\rm tun}^B-V$ characteristics is a sum of two step functions, with two jumps at $\omega = -v_2\Delta k$ and $v_1\Delta k$. The positions of the two jumps provide the information about the edge mode velocities. The differential conductance $G_{\rm tun}^B$ is simply a combination of two $\delta$-functions of $\omega$.

This graphical method can also be used to analyze case (A). Consider the $K=8$ state in the setup Fig. \ref{setup-2} as the simplest example.
For the $K=8$ state, only electron pairs can tunnel through the junction since single electrons are gapped. This does not create much difference for the further analysis. It is convenient
to use bosonization language for the description of the $K=8$ edge. All elementary excitations are bosons with positive momenta $k-k_{F2}>0$ and linear spectrum. Thus, the relation between the momentum and energy remains the same as in the Fermi liquid case.
Hence, the $I_{\rm tun}^A-V$ curve has singularities at $\omega = -v_2\Delta k$ and $\omega = v_3\Delta k$, where $v_3$ is the velocity at the $K=8$
fractional edge. However, the current is no longer a constant above the thresholds because of a different TDOS. We will see below that the current exhibits universal
power-law dependence on the voltage bias near the thresholds.

In the Pfaffian state, case (A) involves three modes: a charged boson mode $\phi_3$ and a neutral Majorana fermion mode $\lambda$ from the upper edge, and the Fermi-liquid mode from the lower edge. They have velocities $v_3$, $v_\lambda$ and $v_2$ respectively.
Any tunneling event involves creation of a Majorana fermion. The spectrum of the Majorana mode is linear: $E=v_\lambda k>0$.
The total energy and momentum of the three modes should be conserved.
As usual, we denote the momentum mismatch between the upper and lower edges as $\Delta k$.
Fig.~\ref{fig:conservation}(b) demonstrates the graphical approach for the Pfaffian state. The left line represents the spectrum of the charged boson
at the upper edge and the right line describes the spectrum of the lower edge. Consider a tunneling process such that an electron from the lower edge tunnels into the upper edge. This may happen at a positive applied voltage. In this process the electron emits a Majorana fermion and creates excitations of the charged boson mode at the upper edge.
The energy and momentum of the electron are the sums of the energies and momenta of the charged boson and Majorana modes.
The unevenly dashed lines of slope $v_\lambda$ in Fig.~\ref{fig:conservation}(b) represent the Majorana fermion.
We will call them $\lambda$-lines. Different $\lambda$-lines start at different {\it occupied} states on the right line and correspond to different momenta of the electron at the lower edge. One can visualize the tunneling process in the following way: an electron with the momentum $k_0$ from the right line slides along the $\lambda$-line (emitting a Majorana fermion with the momentum $k_0-k$) and reaches the left line at $k>k_{F3}$ (otherwise the tunneling is not possible since the momentum change $(k-k_{F3})$ of the Bose mode must be positive). Both energy and momentum are conserved in such picture.
Because the Majorana fermion has a positive momentum the $\lambda$-line points downward and leftward. When $\omega$ is positive and small enough, all the states at the intersections of the left line with the $\lambda$-lines have $k<k_{F3}$, thus, no tunneling happens. At $\omega=v_\lambda\Delta k$, the highest $\lambda$-line intersects the left line at $k=k_{F3}$, so the tunneling becomes possible and contributes a positive current. Thus $\omega=v_\lambda \Delta k$ is the positive threshold voltage. When $\omega$ reaches $v_3\Delta k $, the intersection point of the right and left lines corresponds to $k>k_{F3}$ (an `empty state') at the upper edge
and a filled state at the lower edge. The tunneling process involving those two states and a zero-momentum Majorana fermion becomes possible. This results in another singularity in the $I_{\rm tun}^A-V$ curve. For negative $\omega$, it is expected that a Majorana fermion and an excitation of the charged boson mode combine into an electron and tunnel into the lower edge. The same analysis as above shows that there is no current when $\omega$ is negative and small. When $\omega = -v_2\Delta k$, the tunneling process involving a zero-momentum Majorana fermion becomes possible. Thus, $\omega = -v_2\Delta k$ is the negative threshold voltage in the $I_{\rm tun}^A-V$ curve. We see three singularities in the tunneling current in agreement with the presence of three modes.

For all other proposed fractional states, the graphical method also works but becomes more complicated, so we will not discuss them in detail here. The above discussion, based only on the conservation of energy and momentum, confirms that singularities appear in the $I_{\rm tun}^A-V$ characteristics and they are closely related to the number and nature of the edge modes. In the following section, we discuss the calculations based on the chiral Luttinger liquid theory.

The calculations below involve the velocities of the charged and neutral edge modes. We generally expect charged modes to be faster. Indeed,
in the chiral Luttinger liquid theory the kinetic energy and the Coulomb interaction enter in the same form, quadratic in the Bose-fields.
Since the Coulomb contribution exists only for the charged mode, it is expected to have a greater velocity.

\section{Calculation of the current}
\label{sec:calculations}

We now calculate the tunneling current.
In this section we derive a general expression, valid for all models.
In the next two sections it will be applied to the six models discussed above.

As mentioned above, to the lowest order of the perturbation theory the tunneling current can be separated into three independent parts, $I_{\rm tun} = I_{\rm tun}^A + I_{\rm tun}^B+I_{\rm tun}^C$. The calculation of $I_B$ and $I_C$ is essentially the same. So in the following, we will only consider $I_{\rm tun}^A$ and $I_{\rm tun}^B$.

We will use below the bosonization language which can be conveniently applied to all modes except Majorana fermions.
Thus, we will not explicitly discuss Majorana modes in this section. However, all results can be extended to the situation involving Majorana fermions without any difficulty. Indeed, in the lowest order of the perturbation theory only the two-point correlation function of the Majorana fermion operators is needed. It is the same as for ordinary fermions and the case of ordinary fermions can be easily treated with bosonization.

We consider the Lagrangian density \cite{wen-book04}
\begin{align}
\label{eq9}
\mathcal{L} = & \mathcal{L}_{\rm frac}(t,x) -\frac{1}{4\pi}\partial_x\phi_1(\partial_t + v_1\partial_x)\phi_1\nonumber\\ & -\frac{1}{4\pi}\partial_x\phi_2(-\partial_t + v_2\partial_x)\phi_2-{\cal H}_{\rm tun},
\end{align}
with the tunneling Hamiltonian density
\begin{equation}
\mathcal{H}_{\rm tun} = \sum_n\gamma_A^n\Psi_2^{\dag}(x)\Psi_{\rm frac}^n(x) + \gamma_B\Psi_2^{\dag}(x)\Psi_1(x) + \text{H.c.}. \label{eqn:Htun1},
\end{equation}
where $\Psi_1$ is the electron operator for the integer QHE mode of the upper edge, $\Psi_{\rm frac}^n$ annihilate electrons at the $1/2$-edge,
$\Psi_2$ is the electron operator at the lower edge; Bose-fields
$\phi_j(x)$ $(j = 1, 2)$ represent the right/left-moving integer edge modes of velocities $v_j$ at the upper/lower QHE liquid. The Bose-fields satisfy the commutation relation $[\phi_i(x), \phi_j(x')] = i\sigma_j\pi\delta_{ij}\text{sign}(x-x')$, with $\sigma_1=+1$ and $\sigma_2 = -1$. The Lagrangian density for the fractional QHE edge $\mathcal{L}_{\rm frac}$ depends on the state and will be discussed in detail later.
Eq. (\ref{eq9}) does not include interaction between the inter and fractional QHE modes. Our analysis can be extended to include such interactions
(section V). However, a full analytical calculation of the $I-V$ curve (Sec. VI) is only possible, if it is legitimate to neglect such interactions.


We assume that the line junction is infinitely long and the system is spatially uniform. As discussed above this restricts possible coordinate dependence of the tunneling amplitudes. It will be convenient for us to assume that $\gamma_A^n$ and $\gamma_B$ are independent of the coordinate and absorb the factors
$\exp(-i\Delta k_n x)$ into the electron creation and annihilation operators. The tunneling amplitudes are also assumed to be independent of the applied voltage $V$. In the tunneling Hamiltonian density (\ref{eqn:Htun1}), $\Psi_j(x)$ is the corresponding electron operator of the integer mode $\phi_j(x)$ with $\Psi_j = e^{\sigma_j i\phi_j+ik_{F,j}x}$, where $k_{F, j}$ represents the Fermi momentum. The corresponding electron density
$\rho_j = (\partial_x\phi_j+k_{F,j})/2\pi$. In the fractional edge, there may be several relevant electron operators $\Psi_{\rm frac}^n$. In our calculations, only the most relevant electron operators will be considered, in the sense of the renormalization group theory. Generally, tunneling between integer QHE modes is more relevant than tunneling into the fractional $\nu=1/2$ edge mode. However, as is clear from the above discussion, for weak interactions between integer and fractional modes, the tunneling conductance $G_{\rm tun}^B(\omega)$ is just a combination of two $\delta$-functions. Therefore, the shape of the voltage dependence of the total differential conductance $G_{\rm tun}$ is determined by $G_{\rm tun}^A(\omega)$. Thus, we focus on tunneling into the fractional
channel. In the case of strong interaction, the analysis of the present section has to be slightly modified (Sec. V).

Since the upper and lower edges have different chemical potentials, it is convenient to switch to the interaction representation with $\Psi_{\rm frac}^{n}\rightarrow\Psi_{\rm frac}^{n}e^{-i\mu_1t/\hbar}$, $\Psi_{1}\rightarrow\Psi_{1}e^{-i\mu_1t/\hbar}$ and $\Psi_{2}\rightarrow\Psi_{2}e^{-i\mu_2t/\hbar}$, where $\mu_1 = eV$ and $\mu_2 =0$.
This introduces time-dependence into the tunneling operators (cf. Ref.~\onlinecite{Feldman03}). The electron operator $\Psi_{\rm frac}^n(x)$ can be written in a bosonized form according to the chiral Luttinger liquid theory, $\Psi_{\rm frac}^n(x) = e^{i\sum_I (l_I\phi_I+l_Ik_{F,I}x)}$, or $\lambda(x) e^{i\sum_I (l_I\phi_I + l_I k_{F,I}x)}$, if a Majorana mode $\lambda(x)$  exists.

In order to pay special attention to momentum mismatches, we define
\begin{equation}
\Psi_{\rm frac}^n(x) \equiv \tilde{\Psi}_{\rm frac}^{n}(x) e^{i\sum_I l_I k_{F,I}x}.
\end{equation}
Similar definitions are also made for the integer QHE modes, $\Psi_j(x)=e^{ik_{F,j}x}\tilde{\Psi}_j(x)$.
Thus, the density of the tunneling Hamiltonian can be rewritten in the interaction picture as
\begin{align}
\mathcal{H}_{\rm tun} = &\sum_n\gamma_A^n e^{i\omega t - i\Delta k_{2f}^nx} \tilde{\Psi}_2^{\dag}(x)\tilde{\Psi}_{\rm frac}^n(x)\nonumber\\& + \gamma_B  e^{i\omega t - i\Delta k_{21}x} \tilde{\Psi}_2^{\dag}(x)\tilde{\Psi}_1(x) + \text{H.c.},\label{eqn:Htun}
\end{align}
where $\Delta k^n_{2f} = k_{F,2} - \sum_I l_I k_{F,I}$, $\Delta k_{21} = k_{F,2}- k_{F,1}$ and $\omega = (\mu_2-\mu_1)/\hbar = -eV/\hbar$. It is worth to mention that in the $K=8$ state, electron pairs and not electrons tunnel through the junction, thus in the first term of Eq.~(\ref{eqn:Htun}) $\omega$ should be doubled because the pair charge doubles, and $\tilde{\Psi}_{\rm frac}^{n}$ and $\tilde{\Psi}_2$ should be understood as bosonic operators that annihilate electron pairs.

The operator for the tunneling current density is given by
\begin{equation}
j(t,x) = e\frac{d\rho_2}{dt} =\frac{e}{i\hbar}[\rho_2(x), H_{\rm tun}], \label{eqn:current}
\end{equation}
where $\rho_{2}(x)$ is the electron density of the lower edge, and $H_{\rm tun} = \int\!dx\,\mathcal{H}_{\rm tun}(x)$ is the tunneling Hamiltonian. Expanding the commutator in Eq.~(\ref{eqn:current}) we get
\begin{align}
j(t,x)= \frac{e}{i\hbar}\{ &\sum_n\gamma_A^n e^{i\omega t - i\Delta k_{2f}^nx} \tilde{\Psi}_2^{\dag}(x)\tilde{\Psi}_{\rm frac}^n(x) \nonumber\\&+ \gamma_B  e^{i\omega t - i\Delta k_{21}x} \tilde{\Psi}_2^{\dag}(x)\tilde{\Psi}_1(x) - \text{H.c.}\}. \label{eqn:j(t,x)}
\end{align}

The current can now be calculated with the Keldysh technique. We assume that the tunneling was zero at $t=-\infty$ and then gradually turned on.
Both edges were in their ground states at $t=-\infty$. At zero temperature, the current is given by the expression
\begin{equation}
\label{11-Keldysh}
I_{\rm tun}(t)=\langle 0| S(-\infty,t) I S(t,-\infty)|0\rangle,
\end{equation}
where $\langle 0|$ is the initial state, the operator $I=\int dx j(t,x)$ and
$$S(t,-\infty)={\rm T}\exp(-
i\int^t_{-\infty} H dt'/\hbar)$$
is the evolution operator.
To the lowest order in the tunneling amplitudes, the tunneling current reduces to
\begin{equation}
I_{\rm tun}(t) = -\frac{i}{\hbar}\int dxdx' \int_{-\infty}^t\!\!\!dt' \langle0|[j(t,x), \mathcal{H}_{\rm tun}(t',x')] |0\rangle. \label{eqn:I(t)}
\end{equation}
After a substitution of  Eqs.~(\ref{eqn:Htun}) and (\ref{eqn:j(t,x)}) into Eq.~(\ref{eqn:I(t)}), we can compute the tunneling current since we know all the electron correlation functions from the chiral Luttinger liquid theory.

In the lowest order perturbation theory the current does not contain any cross-terms, proportional to $\gamma_A^i\times(\gamma_A^j)^*$ with $i\ne j$, or $\gamma_A^i\times\gamma_B^*$ .
There are only contributions proportional to $|\gamma_A^i|^2$ or $|\gamma_B|^2$. Thus, without loss of generality we can assume that only one of the tunneling amplitudes is nonzero and write
\begin{equation}
j_{\alpha\beta}(t, x) = \frac{e}{i\hbar}(\gamma e^{i\omega t - i\Delta k x}\tilde{\Psi}_{\alpha}^{\dag}(t, x)\tilde{\Psi}_{\beta}(t, x) - \text{H.c.}).
\end{equation}
The operators $\tilde\Psi_{\alpha}$ and $\tilde\Psi_{\beta}$ represent electron operators on two sides of the junction.
For brevity, we have dropped subscripts of the momentum mismatch $\Delta k$ and tunneling amplitude $\gamma$. Using Eq.~(\ref{eqn:I(t)}), the tunneling current can be expressed as
\begin{align}
I_{\rm tun}^{\alpha\beta} = & -\frac{e|\gamma|^2}{\hbar^2} \int dxdx'\int_{-\infty}^t \!\!dt' (e^{i\omega\Delta t - i\Delta k \Delta x} - \text{c.c.})\nonumber \\ &\times[G_{\alpha\beta}(\Delta t, \Delta x) - G_{\alpha\beta}(-\Delta t, -\Delta x)] \label{eqn:I(t)append}
\end{align}
with $\Delta t = t-t'$, $\Delta x = x-x'$ and
\begin{align}
&G_{\alpha\beta}(\Delta t, \Delta x) \nonumber\\&\quad= \langle0|\tilde{\Psi}_{\alpha}^{\dag}(t, x)\tilde{\Psi}_{\alpha}(t', x')\tilde{\Psi}_{\beta}(t, x)\tilde{\Psi}_{\beta}^{\dag}(t', x') |0\rangle,
\end{align}
and we used the fact that $\langle0|\tilde{\Psi}_{\alpha/\beta}^{\dag}(t,x)\tilde{\Psi}_{\alpha/\beta}(t', x')|0\rangle = \langle0|\tilde{\Psi}_{\alpha/\beta}(t,x)\tilde{\Psi}_{\alpha/\beta}^{\dag}(t', x')|0\rangle$ and the translational invariance for chiral Luttinger liquids. Eq.~(\ref{eqn:I(t)append}) can be simplified as
\begin{equation}
I_{\rm tun}^{\alpha\beta} = -L\frac{e|\gamma|^2}{\hbar^2} \int dyd\tau (e^{i\omega\tau - i\Delta k y} - \text{c.c.})G_{\alpha\beta}(\tau, y), \label{eqn:I(t)append2}
\end{equation}
where $L$ is the length of the junction.

Let there be $N$ right-moving and $M$ left-moving modes in total at both edges. In the chiral Luttinger liquid theory a general expression for the correlation function is
\begin{align}
G_{\alpha\beta}(\tau, y)&=  l_B^2\prod_{i=1}^N \left(\frac{\tau_c}{\delta + i(\tau - y/v_{Ri})}\right)^{g_{Ri}} \nonumber \\&\times \prod_{i=1}^M \left( \frac{\tau_c}{\delta + i(\tau + y/v_{Li})}\right)^{g_{Li}},\label{eqn:correlation function}
\end{align}
where $v_{Ri}$ and $v_{Li}$ denote the velocities of the $i$th right- and left-moving modes, $\tau_c$ is the ultraviolet cutoff and $l_B$ is the magnetic length. This expression relies on the fact that the quadratic Luttinger liquid action can always be diagonalized and represented as the sum of the actions of non-interacting chiral modes.
All the velocities $v_{Ri}/v_{Li}$ and scaling exponents  $g_{Ri}/g_{Li}$ depend on the details of the Hamiltonian and this dependence is discussed separately for each state in Sec.~\ref{sec:results}.
We choose the convention that $v_{R1} < v_{R2}< \cdots < v_{RN}{}$ and $v_{L1} < v_{L2} < \cdots < v_{LM}{}$.
The scaling dimension of the tunneling operator $\tilde{\Psi}_{\alpha}^{\dag}(t, x)\tilde{\Psi}_{\beta}(t, x)$ is $g=1/2(\sum_ig_{Ri}+\sum_ig_{Li})$.

\begin{figure}
\centering
\includegraphics[width = 3in]{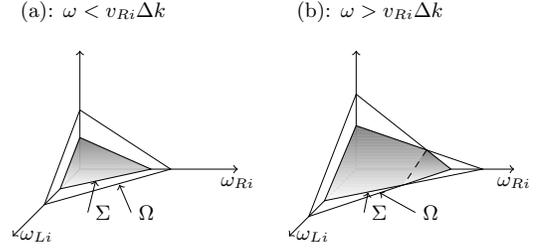}
\caption{A 3-dimensional illustration of the integration volume in the integral (\ref{eqn:I(t)append3}). The integral (\ref{eqn:I(t)append3}) is taken over the volume under the shaded surface in the positive orthant. In panel (a),  $\omega < v_{Ri}\Delta k$ and the $\omega_{Ri}$ axis intersects superplane $\Sigma$ closer to the origin than the plane $\Omega$. In panel (b) $\omega > v_{Ri}\Delta k$ and the order of the intersection points reverses.}\label{fig:planes}
\end{figure}

Using the Fourier transformation
\begin{equation}
\frac{1}{(\delta + it)^g} = \int_{-\infty}^{+\infty}\!\!\!\!\!\!d\omega \, e^{-i\omega t}\frac{|\omega|^{g-1}}{\Gamma(g)}\theta(\omega),
\end{equation}
we integrate out $\tau$ and $y$ in Eq.~(\ref{eqn:I(t)append2}). Then we obtain
\begin{align}
&I_{\rm tun}^{\alpha\beta} = -4\pi^2L\frac{e|\gamma|^2}{\hbar^2} \int[d\omega_{Ri}d\omega_{Li}]\nonumber\\ &\,\times
\Big\{\delta(\omega -\sum \omega_{Ri} - \sum\omega_{Li})\delta(\Delta k - \sum\frac{\omega_{Ri}}{v_{Ri}} + \sum\frac{\omega_{Li}}{v_{Li}})\nonumber \\&  \quad\quad - (\omega \leftrightarrow -\omega, \Delta k \leftrightarrow -\Delta k) \Big\} \nonumber\\ &\,\times\prod |\omega_{Ri}|^{g_{Ri}-1}\frac{\theta(\omega_{Ri})}{\Gamma(g_{Ri})} \prod|\omega_{Li}|^{g_{Li-1}}\frac{\theta(\omega_{Li})}{\Gamma(g_{Li})},
\end{align}
where we absorbed the cutoff $\tau_c$ and the magnetic length $\l_B$ into the tunneling amplitude $\gamma$ for brevity. The two $\delta$-functions represent the energy and momentum conservation. Integrating out $\omega_{R1}$ and $\omega_{L1}$ by using the two $\delta$-functions we obtain our general expression for the tunneling current,
\begin{align}
I_{\rm tun}^{\alpha\beta} = & A \int_0^{\infty} [d\omega_{Ri}d\omega_{Li}]_{i\geq2} \prod_{i\geq2}|\omega_{Ri}|^{g_{Ri}-1}\prod_{i\geq2}|\omega_{Li}|^{g_{Li}-1}\nonumber\\
& \times |\frac{\omega}{v_{R1}} - \Delta k - \sum_{i\geq2}\frac{\omega_{Ri}}{v_{i1}^{RR}} - \sum_{i\geq2}\frac{\omega_{Li}}{v_{i1}^{LR}}|^{g_{L1}-1} \nonumber \\& \times\theta(\frac{\omega}{v_{R1}} - \Delta k - \sum_{i\geq2}\frac{\omega_{Ri}}{v_{i1}^{RR}} - \sum_{i\geq2}\frac{\omega_{Li}}{v_{i1}^{LR}}) \nonumber \\
& \times |\frac{\omega}{v_{L1}} + \Delta k - \sum_{i\geq2}\frac{\omega_{Ri}}{v_{i1}^{RL}} - \sum_{i\geq2}\frac{\omega_{Li}}{v_{i1}^{LL}}|^{g_{R1}-1} \nonumber \\& \times \theta(\frac{\omega}{v_{L1}} + \Delta k - \sum_{i\geq2}\frac{\omega_{Ri}}{v_{i1}^{RL}} - \sum_{i\geq2}\frac{\omega_{Li}}{v_{i1}^{LL}}) \nonumber\\
&- (\omega\leftrightarrow -\omega, \Delta k \leftrightarrow -\Delta k),\label{eqn:I(t)append3}
\end{align}
with
\begin{align}
&A = -L\frac{4\pi^2e|\gamma|^2}{\hbar^2\prod\Gamma(g_{Ri})\Gamma(g_{Li})} (v_{11}^{RL})^{g_{R1}+g_{L1}-1} \\
&v_{i1}^{RR}=\frac{v_{Ri}v_{R1}}{v_{Ri}-v_{R1}}, \quad v_{i1}^{LL}=\frac{v_{Li}v_{L1}}{v_{Li}-v_{L1}}, \quad i\geq2,\\
&v_{i1}^{RL} = \frac{v_{Ri}v_{L1}}{v_{Ri}+v_{L1}}, \quad v_{i1}^{LR} = \frac{v_{Li}v_{R1}}{v_{Li}+v_{R1}}, \quad i\geq1.
\end{align}

Let us discuss the above expression in general before applying  it to the six models. We first consider $\omega>0$.
In that case only the first term in Eq.~(\ref{eqn:I(t)append3}) contributes to $I_{\rm tun}^{\alpha\beta}$. The integration is taken over the volume in the positive orthant of the $(M+N-2)$-dimensional space spanned by $\{\omega_{Ri}, \omega_{Li}\}_{i\geq2}$ under both of the following superplanes
\begin{align}
\Sigma: &\quad\sum_{i\geq2}\frac{\omega_{Ri}}{v_{i1}^{RR}} + \sum_{i\geq2}\frac{\omega_{Li}}{v_{i1}^{LR}} = \frac{\omega}{v_{R1}} - \Delta k ,\\
\Omega: & \quad \sum_{i\geq2}\frac{\omega_{Ri}}{v_{i1}^{RL}} + \sum_{i\geq2}\frac{\omega_{Li}}{v_{i1}^{LL}} = \frac{\omega}{v_{L1}} + \Delta k.
\end{align}
If $\omega < v_{R1}\Delta k$ then the integration volume  is 0 and so is the tunneling current. The tunneling only appears when $\omega > v_{R1}\Delta k$, thus, we see that $v_{R1}\Delta k$ is the positive threshold voltage. It is easy to see that the asymptotic behavior of the tunneling current at $\omega \gtrsim v_{R1}\Delta k$ is
\begin{equation}
I_{\rm tun}^{\alpha\beta}\sim\left(\frac{\omega}{v_{R1}}-\Delta k\right)^{\sum_{i=2}^Ng_{Ri} +\sum_{i=1}^Mg_{Li}-1}.
\end{equation}
Now let us consider the $\omega_{Ri}$-intercepts of the two superplanes, $\Sigma_{Ri}=(\omega/v_{R1}-\Delta k)v_{i1}^{RR}$ and $\Omega_{Ri}=(\omega/v_{L1}+\Delta k)v_{i1}^{RL}$, $i\geq2$. We find that
\begin{align}
\Sigma_{Ri} < \Omega_{Ri},\quad \text{when } \omega < v_{Ri}\Delta k;\nonumber\\
\Sigma_{Ri} > \Omega_{Ri},\quad \text{when } \omega > v_{Ri}\Delta k.
\end{align}
Thus, when $\omega$ passes $v_{Ri}\Delta k$, the shape of the ($M+N-2$)-dimensional integration volume changes, as is illustrated in Fig.~\ref{fig:planes} for the 3D case. This volume change leads to a singularity in the $I_{\rm tun}-V$ curve. The precise nature of the singularities depends on the model and will be discussed in the following section.
For the $\omega_{Li}$-intercepts, $\Sigma_{Li}=(\omega/v_{R1}-\Delta k)v_{i1}^{LR}$ is always smaller than $\Omega_{Li}= (\omega/v_{L1}+\Delta k)v_{i1}^{LL}$, so no extra singularities emerge. Thus, we see that on the positive voltage branch, the tunneling current has $N$ singularities in one to one correspondence with the right-moving modes.

Similar behavior of $I_{\rm tun}^{\alpha\beta}(\omega)$ manifests itself when $\omega <0$,  with singularities at $\omega = -v_{Li}\Delta k$.
Thus, each mode contributes a singularity.

\section{The number of singularities}
\label{sec:V}

\begin{table*}
\centering
\caption{The number of conductance singularities for different models in different setups.}
\begin{tabular}{lcccc}
\hline\hline
    &\hspace{10pt} Boundary of $\nu=5/2$ &\hspace{10pt}  Fig. \ref{fig:setup},
&\hspace{10pt} Fig. \ref{fig:setup}, $\nu=1$ instead of 2,  &\hspace{10pt}Fig. \ref{setup-2}\\
State &       and 2       & \hspace{10pt} strong interaction   & \hspace{10pt} strong interaction & \\
\hline K=8 & 2 & 15 & 8 & 3 \\
$331$      & 6 & 24 & 15 & 8  \\
Pfaffian & $ 3$ & 18 & 10 & 4   \\
Edge-reconstructed Pfaffian & 10 & 61 & 34 & 13\\
Non-equilibrated anti-Pfaffian & 3 & 18 & 10 & 4\\
\hline
\end{tabular}

\label{tbl:table2}
\end{table*}

The analysis of the preceding section allows us to determine the numbers of the conductance singularities in each model for different setups.
Below we consider the $K=8$, $331$, Pfaffian, edge-reconstructed Pfaffian and non-equilibrated anti-Pfaffian states. The special case of the disorder-dominated anti-Pfaffian state will be considered in section \ref{sec:ddap}.

We will need the information about the number of channels and most relevant tunneling operators. This information is discussed in detail in Sec.
\ref{sec:results}. Here we just summarize relevant facts.

We first consider the edge between $\nu=5/2$ and $\nu=2$ states, where only fractional modes exist.
The $K=8$ fractional edge contains a single Bose mode. The 331 edge has two bosonic modes. The Pfaffian edge contains a charged boson and
a neutral Majorana fermion. The edge-reconstructed Pfaffain and non-equilibrated anti-Pfaffian states are characterized by two Bose modes and a Majorana fermion.

The edge between $\nu=5/2$ and $\nu=0$ regions has two additional integer QHE edge modes with opposite spin orientations.

What operators are more relevant depends on the interaction strength as discussed in the next section (see Sec. \ref{sec:neap}). Unless the interaction is very strong, the relative importance of different tunneling operators is the same as in the absence of interaction of different edge modes. Below
we will assume that the set of most relevant operators is the same as for non-interacting modes. Since we consider weak tunneling, only operators which transfer one electron charge will be included. We will have to consider 2-electron operators for the $K=8$ edge between $\nu=5/2$ and $\nu=2$ regions and for the $K=8$ state in the setup Fig. \ref{setup-2} since single-electron tunneling is impossible in those cases.

Thus, the choice of the most relevant tunneling operator into the $K=8$ fractional edge depends on the setup. For the setup Fig. \ref{fig:setup},
the most relevant operator creates an electron pair on the fractional $K=8$ edge and removes an electron from an integer edge channel with the same spin orientation. In the setup Fig. \ref{setup-2}, the most relevant operator transfers an electron pair.

In the 331 state there are two most relevant tunneling operators in the fractional edge. In the bosonization language, both of them are products of exponents of Bose operators
representing two edge channels. The only tunneling operator in the Pfaffian case is the product of a Bose-operator and a Majorana fermion creation/annihilation operator. The reconstructed Pfaffian state has three most relevant tunneling operators. Two of them express via Bose-modes only. The third operator contains also a Majorana fermion. The most important tunneling operator for the non-equilibrated anti-Pfaffian state does not depend on the Majorana fermion.

The above list takes into account only operators that transfer charge into fractional edge modes. In the setup Fig. \ref{fig:setup}, two operators
for the tunneling of spin-up and -down electrons to the integer edge modes must be added. Many more tunneling operators are possible if the integer modes on the edge undergo reconstruction. The reconstruction effects are discussed in the Appendix.

Each tunneling operator contributes two or more singularities into the total conductance. As is clear from the preceding section, the number of the singularities coincides with the number of Bose-modes in the expression for the operator. If the operator contains a Majorana fermion there is an additional singularity. These conclusions are based on the form of the Green function (\ref{eqn:correlation function}). As discussed in the previous section, the expression
(\ref{eqn:correlation function}) can be obtained by diagonalizing the Luttinger liquid Hamiltonian for interacting edge modes. Hence, the number of Bose-modes in the relevant
tunneling operator depends on the details of inter-mode interactions. If all modes interact strongly then {\it after diagonalization}
each tunneling operator contains the same number of Bose modes; this number equals the total number of Bose-channels including all integer QHE channels. If, on the other hand, the interaction between fractional modes and different integer modes is negligible then
the operators of tunneling  into the integer edge modes contain only information about the integer edge channels; the tunneling operators into the fractional modes are independent of the two integer modes on the $5/2$ edge.

We are now in the position to count the singularities in different setups. The results are summarized in Table \ref{tbl:table2}.

Let us first consider tunneling from a single spin-down channel (`spectator' mode) into a boundary between $\nu=5/2$ and $\nu=2$ states
(cf. Ref. \onlinecite{seidel09} for the Pfaffian and non-equilibrated anti-Pfaffian states). There are only two modes (the $K=8$ mode and the `spectator' mode). Hence, there are 2 singularties. For the 331 state, there are 3 modes and 2 tunneling operators. The number of the singularities $2\times 3=6$. The Pfaffian state
is characterized by three modes and one tunneling operator. There are 3 singularities. The reconstructed Pfaffian state has one Majorana mode, two Bose modes plus a `spectator' Bose mode. One tunneling operator expresses in terms of all four modes. The other two tunneling operators do not
contain a Majorana operator. Thus, we find $2\times 3+4=10$ singularities. Finally, the most relevant operator for the non-equilibrated anti-Pfaffian
state does not depend on the Majorana fermion. The remaining three modes result in 3 singularities.

Let us now turn to the setup Fig. \ref{setup-2}. We assume strong interaction between all modes. For the $K=8$ state, we get
$(1 ~{\rm operator})\times (3 ~{\rm modes})=3$ singularities;
for the $331$ state, we get $2\times 4=8$ singularities; for the Pfaffian state, the number of the singularities is $1\times 4=4$; for the reconstructed Pfaffian state we find $2\times 4+5=13$ singularities; the non-equilibrated anti-Pfaffian state is characterized by $1\times 4=4$ singularities.

Next, we consider the setup Fig. \ref{fig:setup}. We first assume that there is no interaction between integer and fractional modes.
With the exception of the $K=8$ state the number of the singularities due to the tunneling into fractional edge channels remains the same as for the
tunneling into the edge between $\nu=5/2$ and 2.
One has, however, to add 4 more singularities due to the tunneling of spin-up and -down electrons into two integer edge channels. Tunneling into
the $K=8$ fractional edge is described by an operator which expresses in terms of three Bose modes. Thus, the total number of the singularities for the
$K=8$ state becomes $3+4=7$.

In the case of strong interaction in the same setup Fig. \ref{fig:setup}, the number of singularities increases.
There are two types of single-electron tunneling operators: tunneling into integer and fractional QHE modes. The first group includes more relevant operators \cite{wen-book04}, cf. Sec. VI.
There are two operators in that group: one for spin-up and one for spin-down electrons. Each of them is responsible for $N$ singularities, where
$N$ is the total number of Bose modes (including 2 `spectator' modes on the lower edge). We will call those singularities `strong'. Thus,
we have $2\times 5=10$  strong singularities for the $K=8$ state; $2\times 6=12$ strong singularities for the $331$ state; $2\times 5=10$
strong singularities for the
Pfaffian state; $2\times 6=12$ strong singularities for the edge-reconstructed Pfaffian state and $2\times 6=12$ strong singularities
for the non-equilibrated anti-Pfaffian state.

Clearly, these numbers alone are not enough to distinguish the states. Additional information comes from
transport singularities due to the next most relevant tunneling operators.
They are responsible for additional `weak' singularities.
In the $K=8$, $331$ and non-equilibrated anti-Pfaffian states
such operators describe tunneling into the fractional modes. Those next most relevant operator were discussed above (see also section VI) and
do not contain Majorana fermions.
Let us find the total number of `weak' and `strong' singularities. In the $K=8$ state we get $10+1\times 5=15$ singularities; in the $331$ state
the answer is $12+2\times 6=24$; and in the non-equilibrated anti-Pfaffian state the answer is $12+1\times 6=18$.

The situation is more complicated
in the Pfaffian and edge-reconstructed Pfaffian states. Just like in the previous three cases we need to take into account tunneling into the fractional edge. This adds $1\times (5+1)=6$ `weak' singularities in the Pfaffian case and $2\times 6 + 7=19$ `weak' singularities for the edge reconstructed state.
There are, however, several additional `weak' singularities for both states. They emerge from tunneling into integer edge channels.

To understand their origin, we need to have a look at the scaling dimensions of the tunneling operators.
Interaction between co-propagating modes has no effect on scaling dimensions of the operators
\cite{wen-book04}. Interaction between counter-propagating modes may change scaling dimensions. Below we will assume
that either 1) all Bose modes are co-propagating or 2) the upper and lower edges in Fig. 1a) are counter-propagating
but the interaction between the two edges is weak. Thus, we will use the same scaling dimensions as for non-interacting modes.

The most relevant operators $\hat T_0$, describing tunneling into
integer edge channels, have scaling dimension 1, Ref. \onlinecite{wen-book04}. The next most relevant operators, describing tunneling into the fractional edge have dimension 2 for both models, Ref. \onlinecite{overbosch08b}. This allows us to calculate how the current scales at low voltages $V$, Ref.
\onlinecite{wen-book04}. We take the square of the renormalized amplitude of the tunneling operator at the energy scale $V$. The renormalized amplitude is  $\sim V^{2d}$, where $d$ is the scaling dimension. Then we
divide it by $V^2$ to reflect the integration over time and coordinate in the expression for the current Eq. (\ref{eqn:I(t)append2}).
 The contribution
of the most relevant operators $I\sim V^{2\times 1-2}=V^0$ and the contribution of the next most relevant operators $I\sim V^{2\times 2-2}=V^2$ in agreement with Section VI.

Now let us consider operators which describe the interaction of the Majorana mode $\lambda$ and an integer QHE Bose-mode $\phi$.
The conservation of the topological charge excludes operators, linear in $\lambda$. Taking into account that $\lambda^2=1$ and that $\phi$ can enter only in the form of a derivative, we find the most relevant interaction term in the action: $Q=\int dx dt \lambda\partial_x\lambda\partial_x\phi$, where
$x$ is the coordinate along the edge. The scaling dimension of the operator $Q$ equals 1. In order to understand the effect of $Q$ on low-energy transport,
let us perform a renormalization group procedure. It should stop at the energy scale $E\sim eV$.
At that scale, different contributions to the current can be obtained from the squares of the renormalized amplitudes of the contributions to the action
describing different tunneling processes
(since the action contains integrations over $t$ and $x$, we will also need to {\it multiply} by $V^2$ to reflect rescaling, cf. Ref. \onlinecite{fsv}).
 At the scale $eV$ the operator $Q$ is suppressed by the prefactor $c\sim eV/\Delta$, where $\Delta$ is the energy gap. The prefactor reflects the scaling dimension of the operator $Q$. Thus, the renormalized action contains the term $cQ$.
Similarly, the contribution to the action, proportional to $T_0$, acquires a prefactor, proportional to $1/V$.

The renormalization group flow generates numerous operators. In particular, the operator $\hat T_1=\hat T_0\lambda\partial_x\lambda$ is generated
from $\hat T_0$ and $Q$. As is clear from the above analysis, it enters the action with the prefactor $\sim c\time 1/V\sim 1$. Hence, its contribution to the current scales as $V^2$ and has the same order of magnitude as for the operators describing tunneling into fractional edges.
This contribution to the current
is singular whenever $eV/\hbar=-\Delta k v_l$, where $\Delta k$ is the momentum mismatch between the integer QHE mode and the `spectator' mode and $v_l$
denote edge mode speeds. The strong singularities due to the operator $\hat T_0$ occur at the same voltages. However, $\hat T_0$ does not contain Majorana
fermions and hence $\hat T_0$ does not generate a singularity at $eV/\hbar=-\Delta k v_M$, where $v_M$ is the speed of the Majorana fermion. On the other hand, $\hat T_1$ contains a Majorana fermion and hence is responsible for an additional `weak' singularity at $eV/\hbar=-\Delta k v_M$. Since there are two
integer edge modes, we discover two additional `weak' singularities.

The above argument completes our discussion of the Pfaffian state. In the edge-reconstructed Pfaffian state there is another mechanism for additional
`weak' singularities. The quadratic part of the action of the fractional edge channels in that state is given by Eq. (\ref{recpfedgeaction}). Let us consider the following four tunneling operators:
\begin{equation}
\hat T_{\uparrow/\downarrow,\pm}= \psi_{d,\uparrow/\downarrow}\psi^\dagger_{u,\uparrow/\downarrow}\lambda\exp(\pm i\phi_n),
\end{equation}
where $\psi_{u/d,\uparrow/\downarrow}$ are annihilation operators for spin-up/down ($\uparrow/\downarrow$) electrons on the upper (u) and lower (d) edges.
$\lambda$ is the Majorana fermion, $\phi_n$ the bosonic neutral mode. The operator $\hat T$ describes electron tunneling between lower and upper integer edge modes. The combination $T'=\lambda\exp(\pm i\phi_n)$ describes charge redistribution between different fractional modes. As is clear from the expressions under Eq. (\ref{recpfedgeaction}), $T'$ is a product of annihilation and creation operators for electrons in fractional channels.
The scaling dimension of the operators $\hat T$ is the same as for the operators describing tunneling into fractional edge modes. Since we have 4 operators and 7 modes, we get 28 additional `weak' singularities.

The total number of `weak' and `strong' singularities is summarized in Table
\ref{tbl:table2}.

A very similar analysis applies to the tunneling between $\nu=5/2$ and $\nu=1$ states. The results are shown in Table \ref{tbl:table2}.

We focused above only on the number of the singularities due to Majorana-fermion and single-Boson excitations at $\omega=v_l\Delta k_n$. All `strong' singularities must be in this class.
All singularities due to the tunneling into fractional edge modes must also be in this class. We were not able to exclude additional `weak' singularities
at $\omega=u_l\Delta k_{\rm integer}$, where $\Delta k_{\rm integer}$ is the momentum mismatch for integer modes and $u_l$ is the speed of a collective excitation. Such singularities might be found if one takes into account contributions to the action, cubic in Bose-fields.
If such additional weak singularities are present it will be easy to separate them from the rest of the singularities. Indeed,
the ratios of all $v_l$ for bosonic modes can be found from the positions of `strong' singularities. Comparison with the positions of `weak' singularities allows then extracting the ratios of all momentum mismatches $\Delta k_m$ and the speed of the Majorana fermion.
After that it is straightforward to check if any singularities due to collective excitations of bosonic modes are present.



The total number of singularities is the  same for the Pfaffian and edge-reconstructed Pfaffian states. However, the number of strong singularities
is different for those models in the setup Fig. \ref{fig:setup} with strong inter-mode interactions.  Thus, the models can be distinguished
just from the number of the singularities in that setup.
At the same time, that number is greater than in other setups and thus requires higher resolution for its detection.
The number of the singularities alone is not enough to distinguish different models in other setups. One also needs information about the nature of the singularities (divergence, cusp or discontinuity of the conductance). The next section discusses the nature of the singularities for the setup
Fig. \ref{fig:setup} with weak interactions and the setup from Ref. \onlinecite{seidel09}.

\section{I-V curves}
\label{sec:results}

In this section we study the setup Fig.~\ref{fig:setup} and focus on the regime of weak interaction with integer QHE modes.
More specifically, we neglect interactions of fractional modes with integer modes (including `spectator' modes on the lower edge)
and the interaction among different integer modes.
Our calculations also apply to the setup Ref. \onlinecite{seidel09}, i.e., tunneling into an edge between $\nu=2$ and $\nu=5/2$ states. In contrast to other cases, the $I-V$ can be analytically computed in the regimes, considered below.

\subsection{Tunneling into integer edge modes.}

Now using the general expression, Eq.~(\ref{eqn:I(t)append3}), we discuss the properties of the tunneling current $I_{\rm tun}$ and
conductance $G_{\rm tun}$ in detail. First, let us consider the simplest case, tunneling between two integer edge modes. Following Eq.~(\ref{eqn:I(t)append3}), it is easy to derive that
\begin{align}
I_{\rm tun}^{B} = & -L \frac{4\pi^2e|\gamma_B|^2v_1v_2}{\hbar^2(v_1+v_2)}\nonumber\\&\times[\theta(\omega-v_1\Delta k_{21})-\theta(-\omega-v_2\Delta k_{21})].
\end{align}
where $v_1$ and $v_2$ are velocities of the upper and lower edge modes respectively, $\Delta k_{21}$ is the momentum mismatch between the two modes. As expected from the qualitative picture, $I_{\rm tun}^B$ is indeed a combination of two step functions and so $G_{\rm tun}^B$ is just a combination of two $\delta$-functions. The two singularities, positive and negative thresholds, appear at $\omega = v_1\Delta k_{21}$ and $-v_2\Delta k_{21}$.

In the following subsections, we will discuss $I_{\rm tun}^A$ and $G_{\rm tun}^A$ as functions of both voltage $\omega$ and momentum mismatch $\Delta k$ for six proposed fractional QHE states.

\subsection{$K = 8$ state}

\begin{figure}
\centering
\includegraphics[width=3.3in]{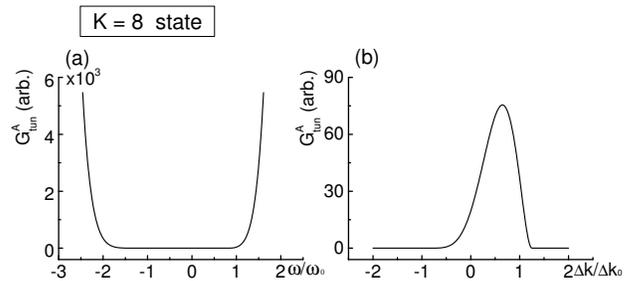}
\caption{(a) Voltage dependence of the differential conductance in the $K=8$ state at a fixed momentum mismatch $\Delta k$ in the case of tunneling into the edge between the states with $\nu=5/2$ and $\nu=2$.  Voltage is shown in units of $\omega_0=v_2\Delta k$, and the conductance is shown in arbitrary units. (b) Momentum mismatch dependence of $G_{\rm tun}^A$ at a fixed voltage. $\Delta k_0$ = $\omega/v_2$. For both curves, we set $v_3/v_2=0.8$.  }
\label{fig:8}
\end{figure}

We distinguish two situations: tunneling into an edge between $\nu=2$ and $\nu=5/2$ states and tunneling into a $5/2$ edge with both integer and fractional modes. We need to distinguish those regimes since they are characterized by different most relevant operators, transfering charge into the
fractional $K=8$ mode.

\subsubsection{Boundary between $\nu=2$ and $\nu=5/2$ states}

In the fractional edge of the $K = 8$ state\cite{overbosch08b}, there is one right-moving boson mode $\phi_3$ with the Lagrangian density
\begin{equation}
\label{k8edge}
\mathcal{L}_{\rm frac} = -\frac{2\hbar}{\pi}\partial_x\phi_3(\partial_t + v_3\partial_x)\phi_3.
\end{equation}
Only electron pairs are allowed to tunnel into the edge.
The electron pair annihilation operator is $\tilde{\Psi}_{\rm frac} = e^{i8\phi_3}$, and the charge density $\rho_{\rm frac} = e\partial_x\phi_3/\pi$. The pair correlation function is $\langle0|\tilde{\Psi}_{\rm frac}^{\dag}(t,x)\tilde{\Psi}_{\rm frac}(0,0)|0\rangle = 1/[\delta + i(t-x/v_3)]^8$, i.e., the scaling exponent $g_3=8$. In the integer edge, the operator $\Psi_2$, Eq. (\ref{eqn:Htun1}), should also be understood as the pair annihilation operator with $\langle0|\tilde{\Psi}_{2}^{\dag}(t,x)\tilde{\Psi}_{2}(0,0)|0\rangle = 1/[\delta + i(t+x/v_2)]^4$ and $g_2=4$. Substituting the scaling exponents and edge velocities into Eq.~(\ref{eqn:I(t)append3}), we
obtain
\begin{align}
I_{\rm tun}^A& = -L\frac{8\pi^2e|\gamma_A|^2}{\hbar^2\Gamma(8)\Gamma(4)} (\frac{v_2v_3}{v_2+v_3})^{11}(\frac{\omega}{v_3}-\Delta k_{2f})^3\nonumber\\&(\frac{\omega}{v_2}+\Delta k_{2f})^7\Big[\theta(\omega -v_3\Delta k_{2f}) - \theta(-\omega - v_2\Delta k_{2f})\Big],
\end{align}
Just like in the case of the tunneling current $I_{\rm tun}^B$ between two integer QHE edges, there are two threshold voltages, the positive threshold $\omega=v_3\Delta k_{2f}$ and the negative one $\omega=-v_2\Delta k_{2f}$. However, in contrast to $I_{\rm tun}^B$, the tunneling current $I_{\rm tun}^A$ increases smoothly as the voltage passes the thresholds.  At $\omega \gtrsim v_3\Delta k_{2f}$, the tunneling current $I_{\rm tun}^A$ behaves as $\sim(\omega-v_3\Delta k_{2f})^3$, and at $\omega\lesssim -v_2\Delta k_{2f}$, $I_{\rm tun}^A\sim(\omega+v_2\Delta k_{2f})^7$. Thus $I_{\rm tun}^A$ follows power laws near the thresholds.  The exponents in the scaling laws for the current near the thresholds provide information about states. However, inter-edge Coulomb interactions may change these exponents and make them non-universal. When $|\omega|\gg v_2\Delta k_{2f}$ and $v_3\Delta k_{2f}$, $I_{\rm tun}^A$ will asymptotically behave like $\sim\omega^{10}$ for both positive and negative voltages. We plotted the differential conductance $G_{\rm tun}^A = \partial I_{\rm tun}^A/\partial \omega$ as a function of $\omega$ at fixed $\Delta k_{2f}$, and a function of $\Delta k_{2f}$ at fixed $\omega$ in Fig.~\ref{fig:8}.

\begin{figure}
\centering
\includegraphics[width=3.3in]{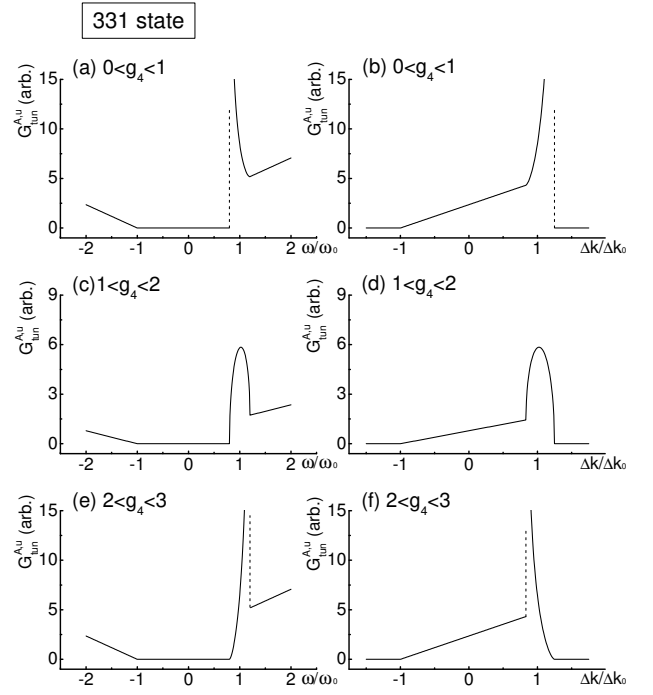}
\caption{Voltage and momentum mismatch dependence of the tunneling differential conductance $G_{\rm tun}^{A, u}$ in the $331$ state; $u$ is either $a$ or $b$.  We have chosen the ratios of the edge velocities to be $v_3/v_2 = 0.8$ and $v_4/v_2=1.2$. The left three panels show the voltage dependence of $G_{\rm tun}^{A, u}$ at a fixed momentum mismatch $\Delta k$ for 3 cases of different scaling exponent ranges: (a) $0<g_4<1$; (c) $1<g_4<2$; (e) $2<g_4<3$; we set $g_4 = 0.5$, $1.5$ and $2.5$ respectively in the plots. Voltage is shown in units of $\omega_0 = v_2 \Delta k$. Panels (b), (d) and (f) show the same three cases for the momentum mismatch dependence of $G_{\rm tun}^{A, u}$ at a fixed $\omega$ with the momentum expressed in units of $\Delta k_0 = \omega/v_2$. The differential conductance is shown in arbitrary units. }
\label{fig:331}
\end{figure}

\subsubsection{Boundary between $\nu=5/2$ and $\nu=0$}
\label{sec:k8-2}

The action remains the same, Eq. (\ref{k8edge}). However, an electron tunneling operator $e^{i8\phi_3-i\phi_1+i\phi_2}$ is present and is more relevant then the pair tunneling operator $e^{i8\phi_3+2i\phi_2}$, considered above. It transfers only one electron into the $5/2$ edge. Two electrons go into the fractional $K=8$ channel and one electron is removed from the spin-up integer channel on the $5/2$ edge.

Our calculations give
\begin{align}
I = &-L\frac{4\pi^2e|\gamma|^2}{\hbar^2 8!}v_{12}\nonumber\\
&\times\left\{\begin{array}{ll}
\vspace{5pt} -v_{23}^{8}(\omega/v_2+\Delta k)^{8}, & \omega < - v_2\Delta k\\
\vspace{5pt}0, & -v_2\Delta k <\omega < v_1\Delta k \\
\vspace{5pt} v_{13}^{8} (\omega/v_1-\Delta k)^{8}, & v_1\Delta k <\omega < v_3 \Delta k \\
v_{23}^{8}(\omega/v_2+\Delta k)^{8}, & \omega > v_3\Delta k
\end{array}
\right.
\end{align}
where $v_{12} = v_1v_2/(v_1+v_2)$, $v_{13}=v_1v_3/|v_3-v_1|$ and $v_{23}=v_2v_3/(v_2+v_3)$. Here we assume $v_3>v_1$.

When $v_1 > v_3$ the tunneling current is
\begin{align}
&I_{tun} = -L \frac{4\pi^2 e |\gamma|^2}{ \hbar^2 8!} v_{23}^8 v_{13} \nonumber\\
& \times\left\{
\begin{array}{ll}
-[(\omega/v_3-\Delta k)^8 & \\
\vspace{5pt}\quad - v_{12}^8/v_{23}^8 (\omega/v_1-\Delta k)^8], & \omega < - v_2\Delta k\\
\vspace{5pt}0, & -v_2\Delta k <\omega < v_3\Delta k \\
\vspace{5pt}(\omega/v_3-\Delta k)^8, & v_3\Delta k <\omega <v_1\Delta k \\
(\omega/v_3-\Delta k)^8 & \quad \\
\vspace{5pt}\quad - v_{12}^8/v_{23}^8 (\omega/v_1-\Delta k)^8, & \omega >v_1\Delta k\\
\end{array}
\right.
\end{align}

In both cases three singularities are found.

\subsection{$331$ state}

The $331$ state \cite{overbosch08b} has the edge Lagrangian density
\begin{align}
\mathcal{L}_{\rm frac} = & -\frac{\hbar}{4\pi}(3\partial_t\phi_{3}\partial_x\phi_{3} -2\partial_t\phi_{3}\partial_x\phi_{4} -2\partial_t\phi_{4}\partial_x\phi_{3} \nonumber\\& +4\partial_t\phi_{4}\partial_x\phi_{4} + \sum_{m,n=3,4}V_{mn}\partial_x\phi_{m}\partial_x\phi_{n}).
\end{align}
Both modes $\phi_3$ and $\phi_4$ are right-moving, and the real symmetric matrix $V$ represents intra-edge interactions. There are two most relevant electron operators in this model, $\tilde{\Psi}_{\rm frac}^a=e^{i3\phi_{3}-i2\phi_{4}}$ and $\tilde{\Psi}_{\rm frac}^b=e^{i\phi_{3}+i2\phi_{4}}$. Before applying Eq.~(\ref{eqn:I(t)append3}) to the calculation of the tunneling current, one needs to compute the correlation functions of $\tilde{\Psi}_{\rm frac}^a$ and $\tilde{\Psi}_{\rm frac}^b$. Since the Lagrangian density $\mathcal{L}_{\rm frac}$ is quadratic, we can rewrite it in terms of two decoupled fields $\tilde\phi_3$ and $\tilde\phi_4$, such that
\begin{equation}
\label{eqn:diagonalied331action}
{\cal L}_{\rm frac} = -\frac{\hbar}{4\pi}\sum_{n=3,4}\partial_x\tilde\phi_n(\partial_t + v_n\partial_x)\partial_x\tilde\phi_n.
\end{equation}
$\tilde\phi_3$ and $\tilde\phi_4$ are linear combinations of $\phi_3$ and $\phi_4$, with $\langle0|\tilde\phi_n(x,t)\tilde\phi_n(0,0)|0\rangle=-\ln [\delta+i(t+x/v_n)]$, where the velocities are
\begin{align}
v_{3,4}=&\frac{1}{16}\Big(4V_{33}+4V_{34}+3V_{44}\nonumber\\
&\mp\sqrt{(1+x^2)}\times|4V_{33}+4V_{34}-V_{44}|\Big),
\end{align}
and $x=2\sqrt{2}(V_{44}+2V_{34})/(4V_{33}+4V_{34}-V_{44})$ is an interaction parameter. Note that $v_3$ is smaller than $v_4$. It is easy to prove that both $v_3$ and $v_4$ are positive, so $\tilde\phi_3$ and $\tilde\phi_4$ are right-moving.  In the limit of strong interaction, $(V_{34})^2\rightarrow V_{33}V_{44}$, $v_3$ approaches 0. The two-point correlation functions of those operators can be expressed as
\begin{align}
&\langle0|\tilde\Psi_{\rm frac}^{u\dagger}(x,t)\tilde\Psi_{\rm frac}^{u}(0,0)|0\rangle \nonumber \\ &\quad\quad=\frac{1}{[\delta+i(t-x/v_3)]^{g_3^u}[\delta+i(t-x/v_4)]^{g_4^u}},\label{Gab-form}
\end{align}
where $u=a,b$ and the scaling exponents
\begin{equation}
g_{3,4}^u=\frac{3}{2}\mp\frac{1-\sigma_u2\sqrt{2}x}{2\sqrt{1+x^2}}\text{sign}(4V_{33}+4V_{34}-V_{44});
\label{eqn:331gx}
\end{equation}
the sign factors $\sigma_a=+1$, $\sigma_b=-1$. It is worth to notice that the sum of $g_3^u$ and $g_4^u$ is always 3.

There are two tunneling operators in the action. They are proportional to $\tilde{\Psi}^a_{\rm frac}$ and $\tilde{\Psi}^b_{\rm frac}$.
These tunneling operators are responsible for two contributions to the current. Based on Eq.(\ref{Gab-form}) and Eq.~(\ref{eqn:I(t)append3}), both contributions have the form
\begin{align}
&I_{\rm tun}^{A,u} = -L \frac{4\pi^2e|\gamma_A|^2}{\hbar^2\Gamma(g_3)\Gamma(g_4)}v_{24}^{g_4}v_{23}^{g_3} (\frac{\omega}{v_2} + \Delta k_{2f})^{2}\nonumber\\&\times\left\{
\begin{array}{ll}
B(1, g_4, g_3), &\omega>v_4\Delta k_{2f} \vspace{5pt}\\
\vspace{5pt}
B(\frac{v_{34}(\omega/v_3-\Delta k_{2f})}{v_{24}(\omega/v_2+\Delta k_{2f})}, g_4, g_3), &v_3\Delta k_{2f}<\omega<v_4\Delta k_{2f}\\
0, &-v_2\Delta k_{2f} <\omega < v_3\Delta k_{2f}\vspace{5pt}\\
-B(1, g_4, g_3),& \omega<-v_2\Delta k_{2f},
\end{array}
\right.
\label{eqn:I,331}
\end{align}
where $u=a$ or $b$. We omitted the index $u$ in the scaling exponents $g_3$ and $g_4$, in the tunneling amplitude $\gamma_A$, and  in the momentum mismatch $\Delta k_{2f}$ in Eq.~(\ref{eqn:I,331}).
$B(z,g_4,g_3)$ is the incomplete Beta function, $v_{23}=v_2v_3/(v_2+v_3)$, $v_{24}=v_2v_4/(v_2+v_4)$ and $v_{34}=v_3v_4/(v_4-v_3)$.

Consider any of the two contributions $I_{\rm tun}^{A,a}$ or $I_{\rm tun}^{A,b}$, Eq.~(\ref{eqn:I,331}). We see expected singularities marked by the edge velocities, with two singularities on the positive voltage side and one on the negative voltage side. The incomplete Beta function $B(z, g_4, g_3)$ has the following asymptotic behaviors
\begin{align}
B(z, g_4, g_3)\sim\left\{
\begin{array}{ll}
z^{g_4}, &\quad z\sim 0\vspace{5pt}\\
(1-z)^{g_3}+\text{const}, &\quad z\sim1
\end{array}
\right.
\end{align}
Thus, when $\omega\gtrsim v_3\Delta k_{2f}$, the differential conductance $G_{\rm tun}^{A,u}\sim(\omega/v_3 - \Delta k_{2f})^{g_4-1}$ is singular at $\omega = v_3\Delta k_{2f}$, if $g_4<1$.  Hence, the differential conductance diverges near the threshold.  Similarly, $G_{\rm tun}^{A,u}$ is singular at $\omega=v_4\Delta k_{2f}$, if $g_3<1$, i.e., $g_4>2$. Hence, the shape of the $G_{\rm tun}^{A,u}\sim\omega$ is quite different  at different values of $g_3$ and $g_4$, i.e., different interaction strengths $x$. Fig.~\ref{fig:331} shows the dependence of $G_{\rm tun}^{A,u}$ on $\omega$ and $\Delta k_{2f}$ in 3 different cases: $g_4<1$, $1<g_4<2$ and $g_4>2$. The total differential conductance $G_{\rm tun}^{A}=G_{\rm tun}^{A,a}+G_{\rm tun}^{A, b}$ has two sets of singularities originating from the two individual contributions to the current. The shape of the curve of $G_{\rm tun}^{A}(\omega)$ depends on the relative values of $\gamma_A^a$ v.s. $\gamma_A^b$, $\Delta k_{2f}^a$ v.s. $\Delta k_{2f}^b$, and $ g_4^a$ v.s. $g_4^b$. Thus, momentum-resolved tunneling allows one to extract considerable information about the details of the edge theory.

\begin{figure}
\includegraphics[width = 3.3 in]{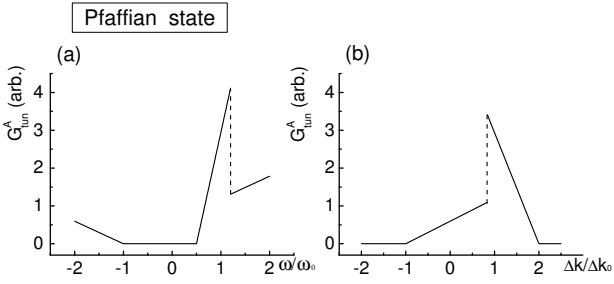}
\caption{(a) Voltage dependence of the tunneling differential conductance $G_{\rm tun}^A$ in the Pfaffian state. The reference voltage $\omega_0=v_2\Delta k$. (b) Momentum mismatch dependence of $G_{\rm tun}^A$ in the Pfaffian state. The reference momentum $\Delta k_0$ = $\omega/v_2$. We set the edge velocity ratios, $v_3/v_2=1.2$ and $v_\lambda/v_2=0.5$. $G_{\rm tun}^A$ is shown in arbitrary units. }\label{fig:Pf}
\end{figure}

\subsection{Pfaffian state}

The Pfaffian state has the edge Lagrangian density \cite{nayak08}
\begin{equation}
\label{pfedgeaction}
\mathcal{L}_{\rm frac} = -\frac{2\hbar}{4\pi}\partial_x\phi_3(\partial_t+v_3\partial_x)\phi_3 + i\lambda(\partial_t + v_{\lambda}\partial_x)\lambda
\end{equation}
where $\phi_3$ is the right-moving charged boson mode and $\lambda$ is the neutral Majorana fermion mode. The most relevant electron operator is $\tilde{\Psi}_{\rm frac}=\lambda\exp(i2\phi_3)$. Its correlation function $G=1/[(\delta+i(t-x/v_3))^2(\delta +i(t-x/v_{\lambda}))]$
equals the product of the correlation function of the Majorana fermion and the correlation function of the exponent of the  Bose-field.
The velocity of the charged mode exceeds the Majorana fermion velocity, $v_{\lambda}<v_{3}$.
A straightforward application of the results of the previous section yields the tunneling current
\begin{align}
&I_{\rm tun}^A = -L\frac{2\pi^2e|\gamma_A|^2}{\hbar^2}v_{2\lambda} \nonumber\\&\quad\times\left\{
\begin{array}{ll}
v_{23}^2(\omega/v_2+\Delta k_{2f})^2, & \omega > v_3\Delta k_{2f}\vspace{5pt}\\
v_{3\lambda }^2(\omega/v_{\lambda}-\Delta k_{2f})^2, & v_{\lambda}\Delta k_{2f}<\omega<v_3\Delta k_{2f} \vspace{5pt}\\
0, & -v_2\Delta k_{2f} <\omega < v_{\lambda}\Delta k_{2f} \vspace{5pt}\\
-v_{23}^2(\omega/v_2+\Delta k_{2f})^2, &\omega <-v_2\Delta k_{2f}
\end{array}
\right.
\end{align}
where $v_{2\lambda}=v_\lambda v_2/(v_2+v_\lambda)$, $v_{23}=v_2v_3/(v_3+v_2)$ and $v_{3\lambda} = v_3v_\lambda/(v_3-v_\lambda)$. Singularities appear again, two of them on the positive voltage side and one on the negative voltage side, quite similar to the results for the $331$ state. However, the Pfaffian state can be distinguished from the $331$ state by a different total number of singularities (Table \ref{tbl:summary}) and
the appearance of a discontinuity for $G_{\rm tun}^A$ at $\omega=v_3\Delta k$ (see Fig.~\ref{fig:Pf}). On the negative voltage side, $G_{\rm tun}^A$ behaves in the same way as in the $331$ state, i.e., it is a linear function of $\omega$.

\subsection{Reconstructed Pfaffian state }

The reconstructed Pfaffian state \cite{overbosch08b} has the Lagrangian density
\begin{align}
\label{recpfedgeaction}
\mathcal{L}_{\rm frac} = &-\frac{\hbar}{4\pi}[2\partial_x\phi_c(\partial_t + v_c \partial_x)\phi_c + \partial_x\phi_n(\partial_t + v_n\partial_x)\phi_n \nonumber\\& +2v_{nc}\partial\phi_c\phi_n] +i\lambda(\partial_t - v_\lambda\partial_x)\lambda,
\end{align}
where $\phi_c$ is a charged mode and $\phi_n$ is a neutral mode. There are three most relevant electron operators $\Psi_{\rm frac}^{\pm} = \exp(i2\phi_c\pm i\phi_n)$ and $\Psi_{\rm frac}^{\lambda}=\lambda\exp(i2\phi_c)$. Thus, we need to consider three tunneling operators, proportional to these three electron operators. As discussed in the previous section they generate three independent contributions to the tunneling current $I_{\rm tun}^A$.
We first discuss the current contributions  which originate from the tunneling terms containing $\Psi_{\rm frac}^{\pm}$. For these two contributions, the situation is quite similar to the $331$ state because the Majorana fermion does not enter the operators $\Psi_{\rm frac}^{\pm}$. We diagonalize the bosonic part of the effective action (\ref{recpfedgeaction}) into the form of Eq.~(\ref{eqn:diagonalied331action}). This requires a transformation from the original fields $\{\phi_c,\phi_n\}$ to two free fields $\{\tilde{\phi}_3,\tilde{\phi}_4\}$ with velocities $\{v_3, v_4\}$ respectively. Then the two-point correlation function $\langle0|\Psi_{\rm frac}^{\pm\dag}(x,t)\Psi_{\rm frac}^{\pm}(, 0,0)|0\rangle$ can be calculated as we did for 331 state. With Eq.~(\ref{eqn:I(t)append3}) we then obtain the same form of the tunneling current $I_{\rm tun}^{A, \pm}$ as in Eq.~(\ref{eqn:I,331}), but with different tunneling amplitudes, momentum mismatches, edge velocities and  scaling exponents. The edge velocities are
\begin{equation}
v_{3,4}=\frac{1}{2}\Big(v_c+v_n\mp(v_c-v_n)\sqrt{1+2x^2}\Big),
\end{equation}
and scaling exponents are
\begin{equation}
g_{3,4}^\sigma =\frac{3}{2}\mp\frac{1+4\sigma x}{2\sqrt{1+2x^2}},
\end{equation}
where $\sigma=+1$ for the case of $\Psi_{\rm frac}^+$ and $\sigma=-1$ for $\Psi_{\rm frac}^{-}$; the interaction parameter $x = v_{nc}/(v_c-v_n)$. It is assumed that $(v_c-v_n)$ is positive. Indeed, we expect the charged mode to be faster than the neutral mode. Thus, for repulsive interactions $x$ is always positive. Similar to the 331 state, different values of $x$ give significantly different shapes of the $G_{\rm tun}^{A,\pm}$ curve, e.g., divergence may appear for certain values of $x$. All  three cases discussed in the subsection on the $331$ state could also emerge in the
edge-reconstructed Pfaffian state.

\begin{figure}
\centering
\includegraphics[width = 3.3in]{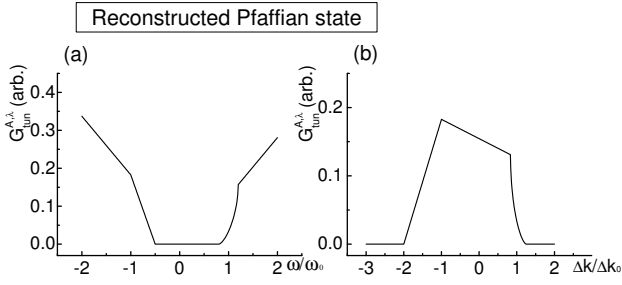}
\caption{The differential conductance $G_{\rm tun}^{A, \lambda}$ in the edge-reconstructed Pfaffian state. Panels (a) and (b) show the voltage and momentum mismatch dependence of $G_{\rm tun}^{A, \lambda}$ (in arbitrary units) respectively. The reference voltage $\omega_0=v_2\Delta k$ and the reference momentum mismatch $\Delta k_0 = \omega /v_2$. We have set $v_\lambda/v_2=0.5$, $v_3/v_2=0.8$, $v_4/v_2=1.2$ and the scaling exponent $g_4=1.5$ } \label{fig:recpf}
\end{figure}

Now let us turn to the tunneling operator, proportional to $\Psi_{\rm frac}^{\lambda}$. In this case
all four modes participate in the tunneling process. The correlation function of the field $\Psi_{\rm frac}^{\lambda}$ is the product of the correlation function of two Majorana fermions and the Bose part. The correlation function for Majorana fermions is the same as for ordinary fermions,
$1/[\delta+i(t+x/v_\lambda)]$. The Bose part has the same structure as in Eq.~(\ref{Gab-form})
with the scaling exponents
\begin{equation}
g_{3,4}^\lambda = 1\mp\frac{1}{\sqrt{1+2x^2}},
\end{equation}
where different signs correspond to indices 3 and 4.
Again, by using Eq.~(\ref{eqn:I(t)append3}) we obtain the following contribution to the tunneling current:
\begin{align}
I_{\rm tun}^{A, \lambda} &=  -L\frac{4\pi^2e|\gamma_A^\lambda|^2}{ \hbar^2\Gamma(g_3+1)\Gamma(g_4)}v_{2\lambda}\,\text{sign}(\omega)\nonumber\\ &
\times\Big[
v_{3\lambda}^{g_3}v_{4\lambda}^{g_4}(\frac{\omega}{v_{\lambda}} +\Delta k_{2f}^\lambda)^2 B(f(\omega), g_4, g_3+1)  \nonumber\\  &\quad- v_{23}^{g_3}v_{24}^{g_4}(\frac{\omega}{v_{2}} +\Delta k_{2f}^\lambda)^2 B(g(\omega), g_4, g_3+1)\Big],
\end{align}
where
\begin{align}
f(\omega)=\left\{
\begin{array}{cl}
\frac{(\omega/v_3-\Delta k_{2f}^\lambda)v_{34}}{(\omega/v_\lambda + \Delta k_{2f}^\lambda)v_{4\lambda}}, & v_3 <\frac{\omega}{\Delta k_{2f}^\lambda} < v_4 \vspace{5pt}\\
1, &\frac{\omega}{\Delta k_{2f}^\lambda} <-v_\lambda \text{ or } >v_4 \vspace{5pt}\\
0, &-v_{\lambda} < \frac{\omega}{\Delta k_{2f}^\lambda} <v_3
\end{array}
\right.
\end{align}
and
\begin{align}
g(\omega)=\left\{
\begin{array}{cl}
\frac{(\omega/v_3-\Delta k_{2f}^\lambda)v_{34}}{(\omega/v_2 + \Delta k_{2f}^\lambda)v_{24}}, & v_3 <\frac{\omega}{\Delta k_{2f}^\lambda} < v_4 \vspace{5pt}\\
1, & \frac{\omega}{\Delta k_{2f}^\lambda} <-v_2 \text{ or } >v_4 \vspace{5pt}\\
0, &-v_{2} < \frac{\omega}{\Delta k_{2f}^\lambda} <v_3
\end{array}
\right.
\end{align}
The dependence of $G_{\rm tun}^{A,\lambda}$ on the voltage $\omega$ and momentum mismatch $\Delta k_{2f}^\lambda$ is illustrated in Fig.~\ref{fig:recpf}. There are no divergencies for any $g_4^\lambda$. All singularities appear as voltage thresholds or discontinuities of the derivative of  $G_{\rm tun}^\lambda(\omega)$. The Majorana fermion mode is responsible for the negative voltage threshold
(we assume that the Majorana is slower than the integer QHE mode at the opposite side of the junction).

Thus, in the edge reconstructed Pfaffian state, three sets of singularities  can be observed. Each set corresponds to one of the three most relevant electron operators. One set contains more singularities than the other two. That extra singularity is due to the neutral Majorana fermion mode.

\begin{figure}
\centering
\includegraphics[width=3.5in]{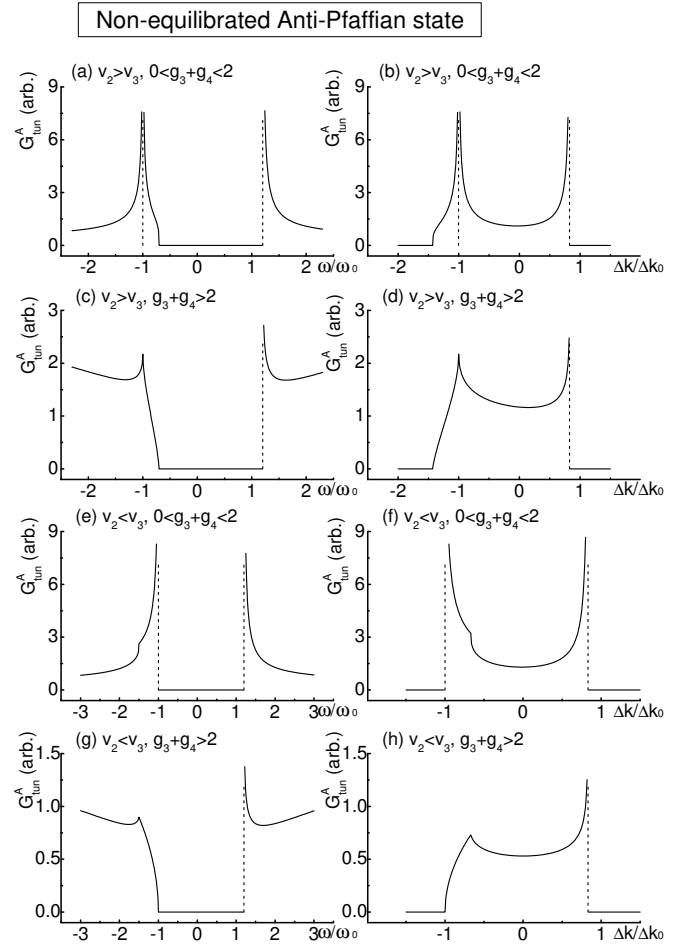}
\caption{Differential conductance $G_{\rm tun}^A$ in the non-equilibrated anti-Pfaffian edge state. All left panels show the voltage dependence of $G_{\rm tun}^A$ and right panels show the momentum mismatch dependence of $G_{\rm tun}^A$, at different choices of $v_3/v_2$ and $g_3+g_4$. In the top four panels, we have chosen $v_3/v_2=0.7$, and in the bottom four panels $v_3/v_2=1.5$. $v_4/v_2=1.2$ for all cases. In panels (a), (b), (e) and (f), illustrating the $0<g_3+g_4<2$ cases, we set $g_3+g_4=1.5$. In panels (c), (d), (g) and (h), illustrating the $g_3+g_4>2$ cases, we set $g_3+g_4=2.5$. The reference voltage $\omega_0=v_2\Delta k$ and the reference momentum mismatch $\Delta k_0=\omega/v_2$. $G_{\rm tun}^A$ is shown in arbitrary units.  }\label{fig:nonApf}
\end{figure}

\subsection{Disorder-dominated anti-Pfaffian state}
\label{sec:ddap}

The very name of this state shows that the momentum-resolved tunneling can only have limited utility in this case.
Indeed, momentum conservation assumes that disorder can be neglected and this assumption fails for the state under consideration \cite{levin07,lee07}. In the disorder-dominated anti-Pfaffian state, the amplitudes of the electron tunneling operators are expected to be random. Thus, one expects that interference between different tunneling sites is irrelevant for the total tunneling current
since the disorder average of the product of two tunneling amplitudes from two different points is zero. Hence, the leading contribution to the current is the same as for the tunneling through a single quantum point contact.
Nevertheless, momentum resolved tunneling might be possible for electron pairs. This happens, if disorder only couples to neutral modes and does not
affect the charged mode. As we see below, the momentum resolved tunneling current of pairs is the same as for the $K=8$ state (Sec. VI B 1).

In the disorder-dominated anti-Pfaffian edge state, there are 3 left-moving  $SO(3)$-symmetric Majorana modes and one right-moving charged mode, with the Lagrangian density \cite{levin07,lee07}
\begin{equation}
\label{eqaction}
\mathcal{L}_{\rm frac} = -\frac{2\hbar}{4\pi}\partial_x \phi_c(\partial_t + v_c\partial_x)\phi_c + i\sum_n^3[\lambda_n(\partial_t-v_\lambda\partial_x)\lambda_n].
\end{equation}
There are three electron operators corresponding to the three Majorana fermions, $\tilde\Psi_{\rm frac}^n=\lambda_ne^{i\phi_c}$, $n=1,2,3$.
Their products yield pair operators. We focus on the pair operator $\exp(2i\phi_c)$ which contains no information about neutral modes.
One can easily verify that its correlation function is the same as the correlation function of the pair operator in the $K=8$ state.
Hence, all results can be taken without modifications from our discussion of the $K=8$ state. Certainly, the total tunneling current includes also
a single-electron part. One may expect that it is greater than the momentum-resolved contribution due to the pair tunneling since the tunneling amplitude is greater for single electrons than for pairs.

\subsection{Non-equilibrated anti-Pfaffian state}
\label{sec:neap}

The non-equilibrated anti-Pfaffian edge has the Lagrangian density \cite{levin07}
\begin{align}
\label{noneqaction}
\mathcal{L}_{\rm frac} = & -\frac{\hbar}{4\pi}[\partial_x\phi_{c1}(\partial_t +v_{c1}\partial_x)\phi_{c1} \nonumber \\ &+ 2\partial_x\phi_{c2}(-\partial_t + v_{c2}\partial_x)\phi_{c2} + 2v_{12}\partial_x\phi_{c1}\partial_x\phi_{c2}] \nonumber \\ & + i\lambda(\partial_t-v_\lambda\partial_x)\lambda.
\end{align}
Again the action can be rewritten in terms of two linear combinations of the Bose fields $\phi_{c1}$ and $\phi_{c2}$: a free left-moving mode $\tilde{\phi}_3$ and a right-moving mode $\tilde{\phi}_4$  with velocities $v_3$ and $v_4$ respectively. From the renormalization group, we find that the most relevant electron operators depend on the interaction strength parameter $x = v_{12}/(v_{c1}+v_{c2})$. Below we will only consider $x<2/3$. The action
(\ref{noneqaction}) is only stable for $x<1/\sqrt{2}$ and hence we ignore a small region $2/3<x<1/\sqrt{2}$ in the parameter space.
For $x<2/3$, the most relevant electron operator is $\Psi_{\rm frac} = e^{i\phi_{c1}}$.

The expression for the tunneling current $I_{\rm tun}^{A}$ and, in particular, the asymptotic behavior near singularities depends on the relative values of $v_2$ and $v_3$, the velocities of the two left-moving modes. If $v_2>v_3$ we obtain the following tunneling current

\begin{widetext}
\begin{align}
I_{\rm tun}^{A} = &-L \frac{4\pi^2 e|\gamma_A|^2}{\hbar^2\Gamma(g_3)\Gamma(g_4)}v_{34}^{g_3+g_4-1}\,\text{sign}(\omega)\nonumber \\ &\times \left\{
\begin{array}{cl}
\frac{v_{24}}{g_3}\Big|\frac{\omega}{v_4}-\Delta k_{2f}\Big|^{g_3} \Big|\frac{\omega}{v_3}+\Delta k_{2f}\Big|^{g_4-1}F(1, 1-g_4, 1+g_3, \frac{v_{24}(\omega/v_4- \Delta k_{2f})}{v_{23}(\omega/v_3 + \Delta k_{2f})}),  & \quad\omega>v_4\Delta k_{2f} \text{ or } \omega<-v_2\Delta k_{2f}\vspace{5pt}\\
\frac{v_{23}}{g_4}\Big|\frac{\omega}{v_4}-\Delta k_{2f}\Big|^{g_3-1} \Big|\frac{\omega}{v_3}+\Delta k_{2f}\Big|^{g_4}F(1, 1-g_3, 1+g_4, \frac{v_{23}(\omega/v_3 + \Delta k_{2f})}{v_{24}(\omega/v_4- \Delta k_{2f})}), & \quad -v_2\Delta k_{2f} < \omega < -v_3\Delta k_{2f} \vspace{5pt}\\
0, & \quad\text{otherwise}
\end{array}
\right.
\end{align}
\end{widetext}
where the scaling exponents equal
\begin{equation}
g_{3,4}= \frac{1}{2\sqrt{1-2x^2}}\mp\frac{1}{2}
\end{equation}
and $F$ is the hypergeometric function.

For the interaction strength we focus on, $0<x<2/3$, we always have $0<g_3<1$ and $1<g_4<2$.  Asymptotically, $I_{\rm tun}^{A}\sim(\omega-v_4\Delta k_{2f})^{g_3}$ when $\omega\gtrsim v_4\Delta k_{2f}$. Thus, $\omega=v_4\Delta k_{2f}$ corresponds to a divergency of the differential conductance. If $\omega\lesssim -v_3\Delta k_{2f}$ then the tunneling current is asymptotically equal to  $(\omega+v_3\Delta k_{2f})^{g_4}$. When $\omega \approx -v_2\Delta k_{2f}$, we have $I_{\rm tun}^{A}\sim (\omega + v_2\Delta k_{2f})^{g_3+g_4-1}$. Hence, when $g_3+g_4<2$, the differential conductance diverges at $-v_2\Delta k_{2f}$, while for $g_3+g_4>2$ only a cusp is present as is shown in Fig.~\ref{fig:nonApf}.

If $v_2<v_3$ then the tunneling current is
\begin{align}
&I_{\rm tun}^{A} = -L\frac{4\pi^2e|\gamma|^2}{\hbar^2\Gamma(g_3)\Gamma(g_4)}v_{23}^{g_3} v_{24}^{g_4}\Big|\frac{\omega}{v_2}+\Delta k_{2f}\Big|^{g_3+g_4-1}\!\!\!\!\!\!\!\text{sign}(\omega)\nonumber \\
&\quad\times\left\{
\begin{array}{ll}
B(\frac{v_{34}(\omega/v_4-\Delta k_{2f})}{v_{23}(\omega/v_2+\Delta k_{2f})}, g_3, g_4), &\frac{ \omega }{\Delta k_{2f}}>v_4 \text{ or } <-v_3\vspace{5pt}\\
B(1, g_3, g_4), & -v_3 <\frac{\omega}{\Delta k_{2f}} < -v_2 \vspace{5pt}\\
0, & \text{otherwise}
\end{array}
\right.
\end{align}
In this case the behavior near $\omega=v_4\Delta k_{2f}$ is the same as above. The behavior near $\omega= -v_3\Delta k_{2f}$ and $\omega = -v_2\Delta k_{2f}$ is also the same as above but these singularities appear now in the opposite order since $v_2<v_3$. The differential conductance is shown in Fig.~\ref{fig:nonApf}.

\section{Discussion}
\label{sec:discussion}

\begin{table*}
\centering
\caption{Summary of singularities in the voltage dependence of the differential conductance $G_{\rm tun}^A$ for different 5/2 states. The ``Modes" column shows the numbers of left- and right-moving modes in the fractional edge, the number in the brackets being the number of Majorana modes. ``A" or ``N" in the next column means Abelian or non-Abelian statistics.  The ``Singularities" shows the number of singularities, including divergencies (S), discontinuities (D) and cusps (C), i.e., discontinuities of the first or higher derivative of the voltage dependence of $G_{\rm tun}^A$.
The table refers to the tunneling into a boundary of $\nu=5/2$ and $\nu=2$ liquids. The case of weak interaction, Fig. 1, is closely related.
 }
\begin{tabular}{lcccc}
\hline\hline
State & \hspace{10pt}Modes\hspace{10pt} & \hspace{10pt}Statistics\hspace{10pt}  &\hspace{10pt}Singularities\hspace{10pt}\\
\hline K=8 & 1R & A &  2C\\
$331$ & 2R & A & 4C+2S or 5C+S \\
Pfaffian & 2R(1) & N & 2C+D \\
Edge-reconstructed Pfaffian & 1L(1) + 2R & N & 8C+2S or 9C+S\\
Non-equilibrated anti-Pfaffian & 2L(1) + 1R & N  & C+2S or 2C+S \\
\hline
\end{tabular}

\label{tbl:summary}
\end{table*}

We have found the number of the transport singularities in different models and setups, Table \ref{tbl:table2}.
We also determined the nature of the singularities for the tunneling into the boundary of the $\nu=5/2$ and $\nu=2$ states, Table \ref{tbl:summary}.
The information from Tables \ref{tbl:table2} and \ref{tbl:summary} allows one to distinguish different models of the 5/2 state.

The results listed in Table \ref{tbl:summary} are also relevant for the transport in the setup Fig. \ref{fig:setup} in the case of weak interactions. Only the case of the $K=8$ state
should be reconsidered as discussed in Sec. \ref{sec:k8-2}. The same types and numbers of singularities will be found in both versions of the setup,
Fig. 1a) and Fig. 1b). In the second case, the control parameter is not voltage bias but the momentum mistmatch between the quantum wire and the
QHE edge.

Certainly, in setup Fig. \ref{fig:setup}, only the total tunneling current
\begin{equation}
\label{cur-disc}
I_{\rm tun} = \sum_iI_{\rm tun}^{A,i}+I_{\rm tun}^B+I_{\rm tun}^C
\end{equation}
and the total tunneling differential conductance $G_{\rm tun}$  can be measured, thus, singularities originating from all three contributions to the current will be seen. Here, $I_{\rm tun}^{B,C}$ describe tunneling into the integer edge modes.
However, these last two contributions to the current (\ref{cur-disc}) always exhibit the same behavior for a weakly interacting system.
They simply give rise to 4 delta-function conductance peaks.


Let us  briefly discuss tunneling between two identical $\nu=5/2$ states. A significant difference from the previous discussion comes from the symmetry of the system. The symmetry considerations yield the identity $I_{\rm tun}(\omega) = -I_{\rm tun}(-\omega)$.
In contrast to our previous discussion, it is no longer possible to read the propagation direction of the modes from the $I-V$ curve as there is no difference between positive and negative voltages.

The tunneling current through a line junction between two $5/2$ states expresses as
\begin{align}
I_{\rm tun} = &I_{\rm tun}^{A}(\Delta k,\omega) + I_{\rm tun}^{A}(-\Delta k,\omega) \nonumber \\ & + I_{\rm tun}^B + I_{\rm tun}^C+ I_{\rm tun}^F,
\end{align}
where $I_{\rm tun}^F$ is the tunneling current between two fractional QHE edges, $I_{\rm tun}^A$ stays for tunneling between integer QHE modes on one side of the junction and fractional QHE modes on the other side of the junction, and $I_{\rm tun}^{B,C}$ describe tunneling between integer QHE modes on different sides of the junction. Since the tunneling operator between two fractional edge modes is less relevant than the other tunneling operators, the contribution $I_{\rm tun}^F$ is smaller than the other contributions. All remaining contributions have already been calculated above.

We considered several different setups. While calculations are similar for all of them, they offer different advantages and disadvantages for a practical realization.
In the setups Fig. ~\ref{fig:setup}, the main contribution to the current comes from the tunneling into integer edge states and additional singularities due to the fractional edge modes are weaker. In the setup shown in Fig. ~\ref{setup-2}, all singularities are due to the tunneling into fractional quantum Hall modes only. However, controlling momentum difference between integer and fractional edges in the setup Fig. ~\ref{setup-2} would require changing the distance between the fractional and integer edge channels. This may potentially result in different patterns of edge reconstruction for different momentum differences and make the interpretation of the transport data difficult. A recent paper \cite{seidel09} considers momentum-resolved tunneling into a 5/2 edge in another related geometry: Electrons tunnel into an edge between $\nu=2$ and $\nu=5/2$ QHE liquids. This allows bypassing the problem of tunneling into integer edge modes. At the same time, it might be more difficult to create a geometrically straight edge in such a setup than on the edge of a sample whereas momentum-resolved tunneling depends on momentum conservation and hence on straight edges. Our results apply to all above setups including that of Ref. ~\onlinecite{seidel09}. In contrast to our paper, Ref.~\onlinecite{seidel09} only considers two candidate states: Pfaffian and non-equilibrated anti-Pfaffian. As discussed above, non-equilibrated anti-Pfaffian state can be probed with a conductance measurement in a bar geometry since its conductance is $7e^2/(2h)$ in contrast to other candidate states. In this paper, we show how the Pfaffian state can be distinguished from several other proposed states which have the same conductance in the bar geometry.

We assumed that the temperature is low. A finite temperature would smear the singularities. To understand the thermal smearing we recall that singularities are obtained at $\hbar\Delta k=|eV/v_m|$, where $v_m$ is an edge mode velocity. A finite temperature can be viewed as a voltage uncertainty of the order of $kT$. Thus, the width of the smeared singularity is $\delta k\sim kT/[\hbar v_m]$. This suggests that the total number of singularities that can be resolved is of the order of $\Delta k/\delta k\sim eV/kT$. The lowest available temperatures in this type of experiments are under 10 mK \cite{gap1}. $eV$ cannot exceed the energy gap for neutral excitations. While there is no data for this gap, it is expected to be lower than the gap for charged excitations.
The latter exceeds 500 mK in high-quality samples \cite{gap2}. This suggests that $N\sim 10$ singularities could be resolved in a state-of-art experiment.
Hence, as the discussion in the Appendix shows, our approach is restricted to the systems with no or only few additional channels due to the reconstruction of the integer edges. Recent observations
of the fractional QHE in graphene \cite{graph1,graph2} may potentially drastically increase relevant energy gaps and the number of singularities that could be resolved.

In conclusion, we considered the electron tunneling into $\nu=5/2$ QHE states through a line junction. Momentum resolved tunneling can distinguish several proposed candidate states. The number of singularities in the $I-V$ curve tells about the number of the modes on the two sides of the junction. The nature and propagation directions of the modes can be read from the details of the $I-V$ curve.

\begin{acknowledgments}
We would like to thank E. Deviatov, Y.~Gefen, M.~Grayson, M.~Heiblum and F.~Li for valuable discussions and comments. CW thanks Weizmann Institute of Science for hospitality. This work was supported by the NSF under Grant No. DMR-0544116 and BSF under Grant No. 2006371.

\end{acknowledgments}

\appendix

\section{Integer edge reconstruction}

In the appendix we determine the number of the conductance singularities in the setup Fig. \ref{fig:setup}
in the presence of additional integer edge modes due to the reconstruction of the integer QHE edge.
As an example, we consider the $331$ state. The situation is similar for other states.

We assume strong interaction between all modes. Additional modes due to edge reconstruction appear in pairs of counter-propagating modes so that the total
Hall conductance is not affected. Let there be $n=(n_\uparrow+n_\downarrow)$ additional modes, where $n_{\uparrow/\downarrow}$ denotes the number of additional modes with the spin pointing up/down. We need to consider two types of operators: 1) most relevant additional tunneling operators
create one electron charge in one of the additional modes; 2)  operators that add one electron charge to one of the integer modes and transfer one electron charge between two other integer modes with the same spin. The operators of the second group are less relevant than the operators of the first group but their contribution to the current can be comparable with the contribution of the operators describing tunneling into fractional modes
(cf. Sec. V).

We find $n$ new operators of the first type. The number of the operators of the second type equals

\begin{eqnarray}
m=(n_\uparrow+1)n_\uparrow(n_\uparrow-1)/2+
(n_\downarrow+1)n_\downarrow(n_\downarrow-1)/2 & \nonumber\\
+(n_\uparrow+1) n_\downarrow(n_\downarrow+1)+(n_\downarrow+1) n_\uparrow(n_\uparrow+1).
\end{eqnarray}
The total number of the modes equals $n+6$. Hence, each tunneling operator  is responsible for $n+6$ singularities and their total number is
 $(4+n+m)(n+6)$. At large $n$ this number grows as $n^4$. Such growth of the number of the singularities limits the utility of the proposed approach
when $n$ is large since it may be difficult to resolve the singularities.


\end{document}